\documentclass[trackchanges]{aastex701}

\usepackage{graphicx} 
\usepackage{natbib} 
\usepackage{amsmath}
\usepackage{amssymb} 
\usepackage{fancyhdr} 
\usepackage{latexsym}
\usepackage{color} 
\usepackage{txfonts}
\definecolor{mypurple}{RGB}{128,0,255}

\graphicspath{{./}}
\usepackage{soul}
\usepackage{epstopdf}
\usepackage{mydefs_iris2_20260127}
\usepackage[normalem]{ulem}
\usepackage{rotating}
\usepackage{booktabs}

\begin{document}

\title{Predicting the thermodynamics in the chromosphere from the translation of SDO data into the IRIS$^{2}$ inversion results using a visual transformer model}
\author[orcid=0000-0002-3234-3070,gname='Alberto', sname='Sainz Dalda']{Alberto Sainz Dalda}
\affiliation{SETI Institute, 339 Bernardo Avenue, Suite 200, Mountain View, CA 94043, USA}
\affiliation{Lockheed Martin Solar \& Astrophysics Laboratory, 3251 Hanover Street, Palo Alto, CA 94304, USA}
\email[show]{asainz.solarphysics@gmail.com} 

\author[orcid=0000-0002-9253-6093,gname='Vishal', sname='Upendran']{Vishal Upendran}
\affiliation{SETI Institute, 339 Bernardo Avenue, Suite 200, Mountain View, CA 94043, USA}
\affiliation{Lockheed Martin Solar \& Astrophysics Laboratory, 3251 Hanover Street, Palo Alto, CA 94304, USA}
\email{vupendran@seti.org} 

\author[orcid=0000-0002-5729-8595, gname='Juno', sname='Kim']{Juno Kim}
\affiliation{Henry M. Gunn Senior High School, 780 Arastradero Rd, Palo Alto, CA 94306, USA}
\email{junokimzone@gmail.com}

\author[orcid=0000-0001-7460-725X,gname='Kyuhyoun', sname='Cho']{Kyuhyoun Cho}
\affiliation{SETI Institute, 339 Bernardo Avenue, Suite 200, Mountain View, CA 94043, USA}
\affiliation{Lockheed Martin Solar \& Astrophysics Laboratory, 3251 Hanover Street, Palo Alto, CA 94304, USA}
\email{kcho@seti.org} 

\author[gname='Paul', sname='S. Killam']{Paul S. Killam}
\affiliation{Lockheed Martin Advanced Technology Center, 3251 Hanover Street, Palo Alto, CA 94304, USA}
\email{paul.s.killam@lmco.com} 

\author[orcid=0000-0003-0975-6659]{Viggo Hansteen}
\affiliation{SETI Institute, 339 Bernardo Avenue, Suite 200, Mountain View, CA 94043, USA}
\affil{Lockheed Martin Solar \& Astrophysics Laboratory, 3251 Hanover Street, Palo Alto, CA 94304, USA}
\email{vhansteen@seti.org} 

\author[0000-0002-8370-952X]{Bart De Pontieu}
\affil{Lockheed Martin Solar \& Astrophysics Laboratory, 3251 Hanover Street, Palo Alto, CA 94304, USA}
\affil{Rosseland Center for Solar Physics, University of Oslo, P.O. Box 1029 Blindern, NO-0315 Oslo, Norway}
\affil{Institute of Theoretical Astrophysics, University of Oslo, P.O. Box 1029 Blindern, NO-0315 Oslo, Norway}
\email{bdp@lmsal.com}



\begin{abstract}

We present \atoi: a visual transformer model that translates a combination of images of the chromosphere and transition region (TR), observed by \aia, and a line-of-sight magnetogram, provided by \hmi, into temperature, line-of-sight velocity (\vlos), velocity of the turbulent motions (\vturb), and electron density (\nne) in the chromosphere. Using the thermodynamic variables obtained from the inversion of the chromospheric lines \mgii\, observed by IRIS, as the target of the model, and the intensity images in the chromosphere and TR, and the photospheric magnetogram as the input, the predicted  T and \nne\ show a strong correlation ($\approx 0.80$) for $\approx 80\%$ of the test inverted data, a moderate-to-strong correlation ($\approx0.63$) for 70\% of the \vturb\ of the target test inverted data, while for the \vlos, the correlation is weak. Therefore, the predicted values by \atoi\ may be used as an estimation of the thermodynamics in the chromosphere, either as a stand-alone result or as complementary information to other chromospheric data observed simultaneously. The execution time employed by \atoi\ to obtain the thermodynamic values in the chromosphere is of the order of a few minutes, being $\le 10$~minutes when using a CPU, and $\le 5$~minutes when using a GPU. \atoi\ opens a new avenue for use of SDO data thanks to the inversions provided by IRIS observables. 

\end{abstract}




\section{Introduction}


Despite the easy observational accessibility of the chromosphere, its complex physical conditions continue to challenge our ability to fully understand the phenomena occurring within it. This layer of the solar atmosphere is characterized by rapid spatial and temporal variations in thermodynamic properties and in the relative strength of the magnetic field, i.e., plasma-$\beta$ (the ratio of plasma pressure to magnetic pressure). Moreover, in the chromosphere, non-local thermodynamic equilibrium (non-LTE) conditions apply. Together, these factors make both the interpretation of observations and the physical modeling of the chromosphere particularly challenging.

From an observational point view, to derive the physics from the observables relies mainly on two approaches: i) the spectroscopic interpretation of the profiles through basic, but meaningful, techniques such as multi-Gaussian fit, center of gravity of the lines, or weak-field approximation for the polarimetric data, or alternately ii) the inversion of the spectral lines \citep{delToroIniesta16}, which in the chromospheric case means to consider both the atmosphere to be under non-local thermodynamic equilibrium conditions, and futhermore the Zeeman effect \citep{Socas-Navarro15,Milic_2018,delaCruzRodriguez19,RuizCobo22}, partial redistribution of the scattered radiation \citep{delaCruzRodriguez16,delaCruzRodriguez19}, and atomic polarization and Hanle effect \citep{Lagg04, AsensioRamos08, AsensioRamos11b, Li_2022}. Other inversion codes in the chromosphere are based on geometrical assumptions for a specific solar feature \citep{Molowny-Horas99,Tziotziou01,Jejcic2022}. While the thermodynamic and magnetic field values obtained by using inversion codes are likely the {\it most accurate} available today, to recover them requires computationally expensive methods and, like any other physical measurement, they carry their own uncertainties. 

Since the first inversion codes started to take advantage of the access to powerful computers, new techniques, nowadays considered as part of the machine learning (ML) and artificial intelligence (AI) environment, were developed to accelerate the process to derive the physical parameters from the spectropolarimetric observations, e.g., \citep{Carroll01c,Socas-Navarro05a,LopezAriste05}. Currently, AI/ML techniques are essential in the analysis of massive amounts of data, and in the inference and interpretation of the information encoded in these data, especially in heliophysics~\citep{Roy_FM, Roy2026}.  Observational astrophysics is not an exception, and indeed is a precursor of the practical use of these techniques. 

In this paper, we develop, test, and evaluate an ML/AI tool that takes advantage of data obtained from two of the most successful NASA missions that observe the Sun: the Interface Region Imaging Spectrograph (IRIS, \citealt{DePontieu14a}) and the Solar Dynamics Observatory (SDO, \citealt{Pesnell12}).
This tool is named \atoi, since it translates the  
intensity passband images obtained by the Atmospheric Imaging Assembly instrument on board SDO (\aia, \citealt{Lemen12}) in combination with the photospheric magnetograms obtained by the Heliosismic and Magnetic Imager (\hmi, \citealt{Hoeksema14}) to thermodynamic parameters in the chromosphere, by using, as a training data set, the same derived parameters from the inversion of the chromospheric lines \mgii\ observed by IRIS and inverted by the inversion tool \irissq\ \citep{SainzDalda19}. 

In Section \ref{sec:data}, we detail how we have built the input datasets and the target datasets used to train \atoi. Section \ref{sec:method} describes the different experiments, i.e., different combinations of variables (or features) considered in the input datasets, the models studied, and the one finally used as the most optimal for these data. A statistical analysis of the correlation between the predicted thermodynamic values obtained by \atoi\ and the ones by \irissq\ on the test dataset, a visual comparison of 3 examples, and an interpretation of the contribution of each feature of the input data, i.e., the \aia\ passband images and the photospheric \hmi\ magnetogram, to the performance of the model are presented in Section \ref{sec:results}. These results are discussed in Section \ref{sec:discussion}. Finally, the conclusions are presented in Section \ref{sec:conclusions}.


\section{Data} \label{sec:data}

We have inverted 1585 IRIS \mgii\ datasets with the \irissq\ inversion tool. For each dataset, we have recast the \aia\ data into IRIS-slit-like rasters, i.e., we \textit{raster} the \aia\ full-disk images at the same location (and time) where the IRIS slit scanned the Sun. Thus, each column of the recast \aia\ data matches the location of the same column of the slit-reconstructed IRIS \mgii\ data. Of course, this construction requires the spatial co-alignment and temporal coincidence between the IRIS slit and the closest-in-time \aia\ data. To this aim, we use the IRIS Slit-Jaw Imager (SJI) to co-align the field of view (FoV) scanned by IRIS with the \aia\ data. In particular, we use the IRIS SJI 1400 $\AA$ data and the \aia\  1600 $\AA$ data for this process. This process involves detecting a common point of reference (keypoints) in both images, then shifting, rotating, and dilating the images. As a result, for any step of the IRIS raster, the closest in time \aia\ images to the IRIS SJI data are co-aligned, and cropped with the closest in time IRIS SJI image. Then, we select the pixels corresponding to the IRIS slit position on these co-aligned, cropped \aia\ images, and we store them in the IRIS-slit-like \aia\ raster. For each step in the raster, we apply the same co-alignment, cropping, and slit-location parameters obtained by the method just described to all the \aia\ channels\footnote{The Python code developed for the construction of the IRIS-slit-like AIA raster can be found at \href{https://github.com/Neontus/mg2hk.git}{https://github.com/Neontus/mg2hk.git}.}. In addition, we use the magnetogram obtained by \hmi\ at the closest time to the time at the middle of the IRIS raster scan. 

\begin{figure*}[h!]
    \centering
    \includegraphics[width=0.59\linewidth]{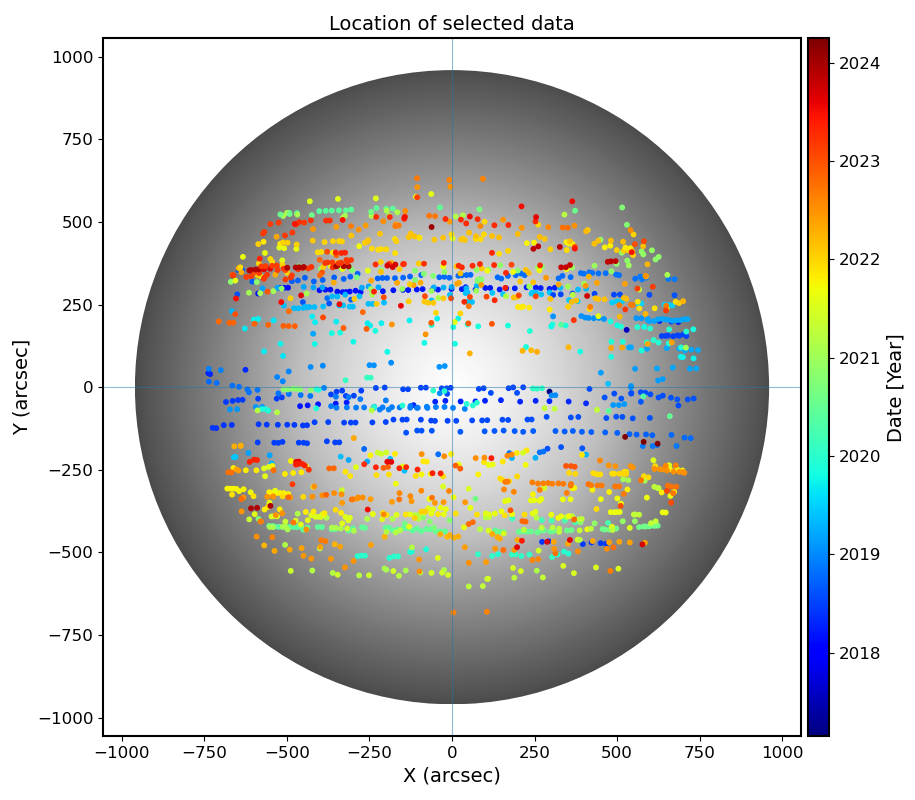}  
    \includegraphics[width=0.39\linewidth]{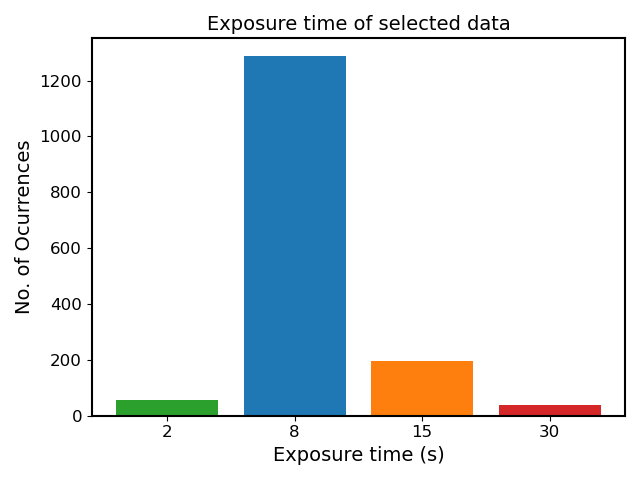} 
    \caption{Left: spatial and temporal distribution of the selected IRIS data on the solar disk. Right: distribution of the exposure time of these data.
    \label{fig:seldata_disk}}
\end{figure*}

The location on the solar disk and the color-encoded date of the selected 1585 IRIS data for this investigation are shown in the left panel of Figure \ref{fig:seldata_disk}. Since the inversion results close to the limb are unreliable, we have not considered data located at a distance from disk center $r>750\arcsec$. The distribution of the exposure time of these data is shown in the right panel of Figure \ref{fig:seldata_disk}. The spectral sampling for all the data is 
51 m\AA, the spatial sampling along the slit is 0\arcsec.33 (2x2 summed), and 0\arcsec.35 perpendicular to the slit.

Thus, for each IRIS dataset considered in this investigation we have the stratified-in-optical-depth ($\tau$) temperature ($T(\tau)$), line-of-sight velocity ($v_{los}(\tau)$), velocity of the turbulent motions or micro-turbulent velocity ($v_{turb}(\tau)$), and electron density ($n_e(\tau)$), the intensity images at the \aia passbands at 171 \AA, 193 \AA, 304 \AA, 1600 \AA, and 1700 \AA, and the photospheric magnetogram obtained by \hmi\ from observing the \ion{Fe}{1} 6173 $\AA$ line. Since the optical depths where the \mgii\ lines are sensitive to changes in the thermodynamics correspond to the chromosphere, we consider the thermodynamic values averaged in an optical depth interval of $\pm0.2$ around \ltau\ at -5.2 (high chromosphere), -4.2 (mid chromosphere), and -3.2 (temperature minimum). Summarizing, for each IRIS dataset, we have the following co-aligned, quasi-simultaneous data: 12 thermodynamic maps (4 variables evaluated at 3 chromospheric optical depths), 5 intensity images corresponding to the chromosphere, transition region, and corona, and 1 photospheric magnetogram. 

\section{Methodology}\label{sec:method}

We use the thermodynamic maps obtained by \irissq, the \aia\ intensity images, 
and the \hmi\ magnetograms to train a visual transformer~\citep[ViT][]{vit_2020} model to predict the
thermodynamic values in the chromosphere from the \aia\ images and the \hmi\ magnetograms, or, 
using the machine learning jargon, to translate the intensity images from the chromosphere,
TR, and corona, and the photospheric line-of-sight component of the vector magnetic
field into the thermodynamics in the chromosphere. 

We have run several experiments and used different metrics to evaluate their impact on the final prediction. Thus, we have selected a combination of features, i.e., the values on the maps from the AIA channels and the HMI magnetograms, that provides the highest correlation to the \irissq\ thermodynamic results. These experiments may be summarized as follows:

\begin{itemize}
    \item {\it Individual channel} experiments: the ViT model is trained considering
    data from only one AIA channel.
    \item {\it Lower atmosphere channels} experiments: these include a combination of two or more of the AIA 1600, AIA 1700, and AIA 304 channels.
    \item {\it All AIA} experiment: All the AIA channels are simultaneously considered in the input of the model.
    \item {\it Lower atmosphere with the photospheric magnetogram} experiments: These experiments add the HMI magnetogram to the channels considered as input in the {\it Lower atmosphere channels} experiments.
    \item {\it All AIA with the photospheric magnetogram} experiment: Similar to the {\it All AIA} experiment, now including the HMI magnetogram.
\end{itemize}

An essential ingredient of models like ViT is the {\it position encoding}. This input provides the contextual spatial position of each pixel to the model, enabling it to learn representation of data both locally and globally. Thus, in addition to the AIA and HMI inputs, we also include the location of the pixel to provide the model with additional information about the relative position of each pixel. To this end, we use fixed position encoding of the sinusoidal form~\citep{posenc_2017,transformer_vaswani}. 

For each pixel in the image $\mathbb{P}_{i,j}$ at the location $(i,j)$ on the image, the pixel location normalized between $[-1,1]$, such that $i = -1$ corresponds to the left edge of the image, and so on. For each position, we compute the position encoding $\mathrm{PE}_{i,j,\omega}$ as: $$\mathrm{PE}_{i,j,\omega} = [\sin(2i\pi\omega), \sin(2j\pi\omega),\cos(2i\pi\omega), \cos(2j\pi\omega)].$$  We consider the set of $\omega$ as integers from $0$ to $6$. This gives us $24$ extra inputs, apart from the physics variables.

In addition to these experiments, we have created a deep neural network (DNN) to predict the \irissq\ thermodynamics, considering all the AIA channels and the HMI magnetogram as input simultaneously. Due to the simplicity of this model and its acceptable, but not optimal performance, we consider this experiment as the baseline of our study. The baseline DNN is defined as a pixel-to-pixel translator, which does not have any local or global context. This model takes in the vector at each pixel consisting of all AIA data, the photospheric magnetogram, and the position encoding vector as input, and predicts the \irissq\ thermodynamics. The DNN architecture is summarized in Table.~\ref{tab:dnn_arch}. The model is trained to minimize the $\mathrm{L}_1$ distance between the target and prediction. For a target of $\mathrm{y}$ and a prediction of $\hat{\mathrm{y}}$, we have:
\begin{equation}
    \mathrm{L}_1 = \frac{1}{N}\sum_{i}^{N}|\mathrm{y} - \hat{\mathrm{y}}|.
    \label{eqn:l1loss}
\end{equation}
\begin{table}[h!]
\centering
\begin{tabular}{l l}
\hline
\textbf{Layer} & \textbf{Configuration} \\
\hline
Linear 1 & $30 \rightarrow 128$ \\
Dropout & $p = 0.3$ \\
ELU & activation \\
Linear 2 & $128 \rightarrow 64$ \\
Dropout & $p = 0.3$ \\
ELU & activation \\
Linear 3 & $64 \rightarrow 32$ \\
Dropout & $p = 0.3$ \\
ELU & activation \\
Output Linear & $32 \rightarrow 12$ \\
\hline
\end{tabular}
\caption{Architecture of the baseline DNN.}
\label{tab:dnn_arch}
\end{table}
Now, we introduce the model developed for this investigation: \atoi. 

\subsection{The \atoi\ visual transformer model}\label{sec:model}

Our main working model developed here is the \atoi. The model is an Image-to-Image vision transformer~\citep{transformer_vaswani, vit_2020}. Consider the input image of size  $x\in\mathbb{R}^{C_{in}\times H\times W}$, where $C_{in}$ is the number of input channels (=30, if including all the AIA passbands (5), HMI image (1), and the position encoding parameters (24)), $H, W$ are the height and width of the image respectively. This image is split into non-overlapping patches of size $P$, such that each patch is $x_P\in\mathbb{R}^{C_{in}\times P\times P}$, giving $HW/P^2$ patches per image. Each patch is flattened, and projected linearly onto a latent dimension $D$, turning the patches into \emph{tokens} ($\mathbb{E}$). Each token is encoded with a stack of transformer encoder layers to an embedding space. We use the standard implementation of the transformer encoder from Pytorch~\citep{Paszke2019PyTorch}, and for a detailed description we refer to ~\cite{transformer_vaswani, layernorm}. In short, the transformer encoder computes multi-head self-attention with scaled dot product attention, a layer-normalization, and a feed forward linear projection which are combined with the input with residual connections. This generates a representation of the input data that has both local and global contexts (Memory $\mathbb{M}$). The transformer decoder layer considers both $\mathbb{E}$ and $\mathbb{M}$, and produces the \irissq\ thermodynamic parameters. The model architecture details are given in Table.~\ref{tab:vit_arch}.

\begin{table}[h!]
\centering
\begin{tabular}{l l}
\hline
\textbf{Parameter} & \textbf{Value} \\
\hline
Patch size & $8$ \\
Tokenizer & $8\times8\times C_{in} \rightarrow 256$ \\
Activation & RELU \\
Encoder: Number of heads & $8$ \\
Encoder: Feed forward Network size & $512$ \\
Encoder: Number of layers & $8$ \\
Decoder: Number of heads & $8$ \\
Decoder: Feed forward Network size & $512$ \\
Decoder: Number of layers & $8$ \\
\hline
\end{tabular}
\caption{Architecture of the \atoi\ model.}
\label{tab:vit_arch}
\end{table}
\textbf{Loss function}: Given a target of $\mathrm{y}$ and a prediction of $\hat{\mathrm{y}}$, the \atoi\ model is trained on: 
\begin{equation}
    \mathrm{L}_{MS} = \mathrm{L}_1(\mathrm{y},\hat{\mathrm{y}}) + \alpha\frac{1}{K}\sum_{k}^{K}\mathrm{L}_1\left(\vec{\nabla}\mathcal{A}_k[\mathrm{y}],\vec{\nabla}\mathcal{A}_k[\hat{\mathrm{y}}]\right).
    \label{eqn:vit}
\end{equation}
Eq.~\ref{eqn:vit} is a combination of two terms: the standard pixel-wise L1-loss term, and a loss term that minimizes the distance between the spatial gradients ($\vec{\nabla}$) of the target and prediction. Pixel-wise L1 loss enables a correct reconstruction of each pixel, while the spatial gradient term helps match the spatial coherence between the prediction and the target. To ensure coherence matching at different spatial scales, the gradients are computed on images which are binned spatially ($\mathcal{A}_k$), such that $k$ is the number of pixels over which the binning is performed. The spatial gradient distance is computed along the width and height for multiple $k$ values in an L1 sense, and these are all averaged by the total number of averaging scales. In this work, we consider $k = [1,2,4]$, which correspond to no binning, binning by $2\times2$ and $4\times4$ pixels respectively. The $\alpha$ is set to $0.2$ after a parameter search. 

\subsection{Data preparation and model training}
The data are split sequentially in time in a proportion of $80:10:10$ as training, validation, and testing sets. This split has not been done randomly. Thus, the training set considers the first 80\% of the selected data. We proceed in this way to avoid the chance of having two rasters observed consecutively on the same FoV, being one alocated in the training set, and the other either in the validation or the test set. We have excluded any data that includes pixels affected by the South Atlantic Anomaly (SAA) from the training set.

Each input and target is standardized per channel by subtracting the mean and dividing by the standard deviation of the whole dataset. The images are zero-padded to a size of $[296,192]$, to ensure divisibility by the selected patch size. The models are trained with a learning rate of $10^{-4}$ using the Adam optimizer~\citep{kingma2014adam}.
\subsection{Variable input size to the model}
The core idea behind the development of this tool is to ingest data of variable spatial dimensions, and present a plausible set \irissq\ thermodynamic parameters. In the \atoi\ model, the inputs have pre-defined spatial sizes. To this end, for an image of any given size, we zero-pad the boundaries and extract rolling-patches the size of our predefined input size. The estimated inversions are then put together, while the pixels with overlapping predictions are averaged out. This enables \atoi\ to be a general purpose tool, and not bound by specific image sizes. 

\section{Results}\label{sec:results}

To quantify the quality of the results obtained by the prediction tool presented in this paper, we have used 3 different approaches. 

First, we have analyzed the linear correlation between the thermodynamics values obtained by \irissq\ (or simply, \irissq\ thermodynamics) and the ones obtained by \atoi. In this way, we can compare one-to-one the results obtained by both methods.
We have done this comparison for all the selected data in the test data set, for 11 experiments considering a combination of AIA channels and the HMI magnetograms as the features of the input of \atoi. This study allows us to conclude which experiment is the best in terms of correlation.

To gain a deeper understanding of the contribution of each feature to the prediction of  \atoi, we investigate the {\it Shapley values method}. These values indicates how much each feature changes the prediction with respect to the mean.

Finally, we show a visual comparison of the \irissq\ thermodynamics and the \atoi\ at the selected optical depths for 3 datasets of the test data set, which have been selected to show different issues that the results from \atoi\ may present.

\subsection{Statistical comparison}

We have used a test data set consisting of 185 datasets. None of these datasets were used to train the model. For each dataset, we use the trained \atoi\ model to predict the thermodynamic values for the experiments described in Section \ref{sec:method}. Then, we have calculated the linear correlation for each thermodynamic variable at the selected optical depth intervals obtained in these experiments with the ones obtained by \irissq\ for the testing set. The means of the Pearson correlation coefficients, $<r>$, for these comparisons for each experiment are detailed in Table \ref{table:all_lincorr_original}, while Table \ref{table:all_lincorr_Zscore} shows the same comparison when the outliers for values above and below $\pm3\times\sigma$ are removed, with $\sigma$ being the standard deviation, i.e., z-score=$\pm$3.
The error in the mean of the Pearson correlation in these tables is the standard deviation of the $r$ corresponding to all the datasets in the test data set. In this study, we consider a linear correlation between two variables is strong when $1.0<|r|<0.8$, moderate when $0.8\le|r|\le0.5$, and weak when $0.5\le|r|\le0.0$. It is important to clarify that in our study we have only considered a linear correlation between the target variable (the \irissq\ thermodynamic values), and the predicted variable (the thermodynamic values obtained by \atoi). Therefore, a weak linear correlation means that there is no linear correlation between these variables, but that does not necessarily indicate that there is no other kind of correlation. 
For the sake of simplicity, in this paper, we have omitted the term ``linear'' when we talk about the strength of the correlation between these variables, e.g.: ``strong correlation'' refers, in this paper, to ``strong linear correlation''. 

A few conclusions can be derived from the inspection of Table \ref{table:all_lincorr_original}: i) using individual coronal or TR channels (AIA 193, AIA 171, and AIA 304) as the input to the model shows a weak correlation, ii) using individually or together the channels corresponding to the lower part of the solar atmosphere (AIA 1700 and AIA 1600) shows a moderate linear correlation, with the contribution of adding AIA 304 small, being only slightly relevant for $v_{turb}$, iii) considering all the AIA chanels together is similar to considering just AIA 1700, AIA 1600, and AIA 304 together, iv) including the HMI magnetogram does not increase significantly the linear correlation.

When the outliers are removed (shown in Table \ref{table:all_lincorr_Zscore}), we see that the correlation coefficient $r$ typically increases its value by $\approx0.1$, becoming closer to a strong correlation for T and \nne, and a moderate-strong correlation for $v_{turb}$.

Both tables show that the worst correlation occurs for the line of sight velocity $v_{los}$, being weak at any optical depth for all the experiments. The stronger correlations for T, \vturb, and \nne\ occur for the experiments {\it ``Lower atmosphere with the photospheric magnetogram''} and {\it ``All AIA with the photospheric magnetogram''}, with almost identical $r$ values, being the strongest correlation for \nne\ at \ltau=-5.2, followed very closely by the temperature T at \ltau=-4.2.

The linear correlation coefficients of all the experiments are larger than the ones of baseline experiment, which has a moderate correlation - in the low end of the range - for T and \nne, and a weak correlation for \vturb\ and \vlos. 

\begin{figure*}
    \centering
    \includegraphics[width=1\linewidth]{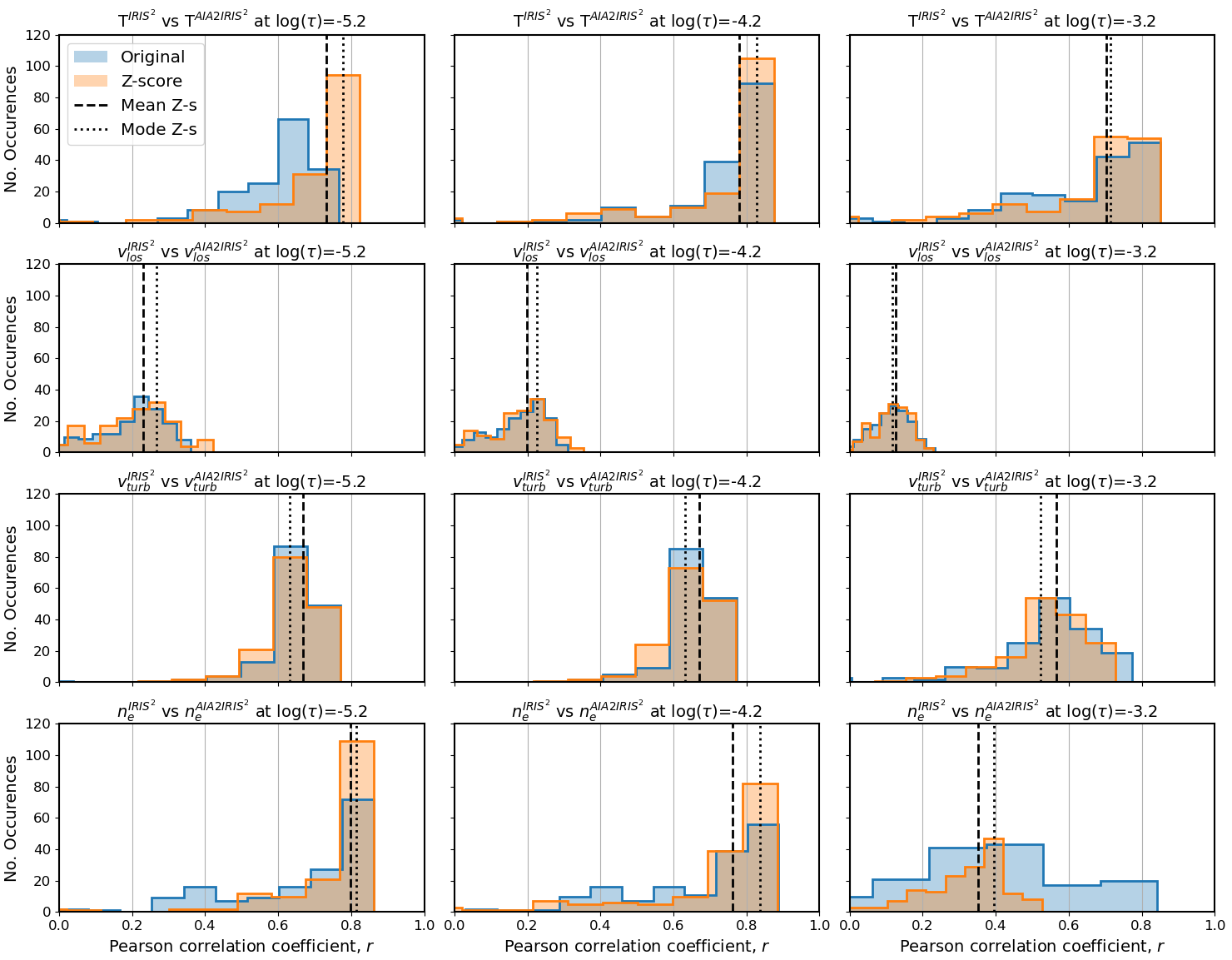}
    \caption{Distributions of the Pearson correlation coefficients ($r$) from the linear correlation between each thermodynamic variable (rows) obtained by \irissq\ and the corresponding one obtained by \atoi\, at the selected optical depths (columns), for the original data (blue) and the Z-score filtered data (orange) of the experiment {\it ``1600 1700 304 HMI''}. The mean and the mode of each distribution for the Z-score filtered data is indicated by dashed and dotted lines, respectively.}
     \label{fig:histograms}
\end{figure*}

The values shown in Tables \ref{table:all_lincorr_original} and \ref{table:all_lincorr_Zscore} 
tell us about the mean value of the linear correlation coefficients for the test data set. For a 
deeper description of the behavior of the correlations between the \irissq\ thermodynamic values and the ones predicted by \atoi, Table \ref{table:distribution_mean} shows $<r>$ for the linear correlation when the outliers have been removed, i.e., the same $<r>$ as in Table \ref{table:all_lincorr_Zscore}, but we now include (in parentheses)
the percentage of datasets for which the correlation coefficient falls within the interval $<r>\pm0.1$. Similarly, Table \ref{table:distribution_mode} shows the mode, $m$, of the distribution of $r$ when the outliers have been removed. Again, the number in parenthesis in this table shows the percentage of the test data set within the interval of $m-0.1 < m \le m+0.1$. As an example, Figure \ref{fig:histograms} shows  the distribution of $r$ for the experiment {\it``1600 1700 304 HMI''} for the original data (blue-shaded bars), the Z-score filtered data (orange-shaded bars), and the mean (dashed) and the mode (dotted line) of the latter.

If we compare these tables, we see that the mode values as slightly higher than the mean values in T and \nne, and slightly smaller for \vturb. For \vlos, the values of the mode show a weak correlation, just like for the mean. Another noticeable finding in this table is the larger percentage of data around the mode than around the mean, with $\approx80\%$ of the data in the test data set showing a mode of $r$ of $\approx0.8$ for T and \nne\ at the optical depths \ltau=-5.2 and \ltau=-4.2.
However, for \vturb, a larger percentage of data is located around the mean, instead of around the mode, with $\approx82\%$ of these data having an $r\approx0.65\pm0.10$ at the the optical depths \ltau=-5.2 and \ltau=-4.2. In T and \vturb, the mean and the mode at \ltau=-3.2 is slightly smaller than at \ltau=-5.2 and \ltau=-4.2, but still showing a moderate correlation, while for the \nne\ shows a weak correlation at optical depth \ltau=-3.2. The maximum mode and its percentage of data in the interval $m-0.1 < m \le m+0.1$ for each thermodynamic variable at any optical depth is shown in bold text in Table \ref{table:distribution_mode}. 

An important conclusion from Tables \ref{table:distribution_mean} and \ref{table:distribution_mode} is that for any thermodynamic variable, at any optical depth, the behavior of the mean and the mode - and the percentages of data of the test data set within the intervals previously defined - are very similar for any experiment considering simultaneously AIA 1600, 1700, and 304 as features of the input to \atoi. In other words, the values along a variable:optical-depth sub-column from the 6$^{th}$ row to the 11$^{th}$ are very similar. A little bit more variation in the mode occurs along the \ltau=-3.2 column for all the thermodynamic variables. 

One interesting fact to note that including all AIA channels ({\it ``All AIA''}) for \ltau=-3.2 appears to increase the mode of the correlation coefficient and increase the percentages associated with the mode to 77\%, compared to experiments that only use the lower atmosphere information (40-50\%). At the same time, including all AIA channels appears to slightly decrease the mode of the correlation coefficient for \vturb\ at \ltau=-3.2, and decrease the percentages to 45\%, compared to 70 to 80\% for experiments like {\it ``1600 1700''} and {\it ``1600 1700 304''}. It thus appears that including information from the higher atmosphere has a positive effect on predicting the temperature in the temperature minimum region and negative effect on predicting the turbulence in the temperature minimum region.  
It seems that including the AIA channels corresponding to the higher solar atmosphere has an impact in T and \vturb\ in the lower atmosphere, making the percentage of data around the mode different for these variables at that optical depth. It is not straightforward to interpret this result. The temperature minimum region is typically not very sensitive to conditions in the transition region or corona. One exception may be that in mossy plage the high coronal pressure squeezes the transition region to lower levels so that transition region and coronal intensities may have some correlation with temperature minimum conditions. However, it is also true that \irissq\ has an in-built problem with resolving the degeneracy between temperature and microturbulence. So perhaps our finding is related to a bias in the chromospheric inversions that depends on the type of region studied, e.g., mossy plage may have a different bias towards temperature vs. microturbulence than other regions. This may be related to the underlying inversions and how they find local vs. global minimima in their quest to find representative atmosphere to reproduce the observed Mg II spectra. A more detailed investigation is required that focuses on the limitations of chromospheric inversions used here, but is not within the scope of the current work.   In the following section, we will investigate the contribution of each feature in the performance of the model, and will propose an explanation for the effects we described in the above of this paragraph.

\begin{sidewaystable}
\begin{tabular}{c *{12}{c}}
\toprule
\multicolumn{1}{c}{Experiment} &
\multicolumn{3}{c}{T} &
\multicolumn{3}{c}{$v_{los}$} &
\multicolumn{3}{c}{$v_{turb}$} &
\multicolumn{3}{c}{$n_{e}$} \\
\cmidrule(lr){2-4} \cmidrule(lr){5-7} \cmidrule(lr){8-10} \cmidrule(lr){11-13}

&
-5.2 & -4.2 & -3.2 &
-5.2 & -4.2 & -3.2 &
-5.2 & -4.2 & -3.2 &
-5.2 & -4.2 & -3.2 \\ 
\cmidrule(lr){2-2} \cmidrule(lr){3-3} \cmidrule(lr){4-4}
\cmidrule(lr){5-5} \cmidrule(lr){6-6} \cmidrule(lr){7-7}
\cmidrule(lr){8-8} \cmidrule(lr){9-9} \cmidrule(lr){10-10}
\cmidrule(lr){11-11}\cmidrule(lr){12-12}\cmidrule(lr){13-13}


 193  & 0.34$\pm$0.14 & 0.44$\pm$0.17 & 0.33$\pm$0.12 & 0.10$\pm$0.07 & 0.08$\pm$0.06 & 0.02$\pm$0.03 & 0.11$\pm$0.10 & 0.12$\pm$0.11 & 0.32$\pm$0.14 & 0.41$\pm$0.16 & 0.42$\pm$0.15 & 0.30$\pm$0.38 \\

 171  & 0.32$\pm$0.13 & 0.39$\pm$0.15 & 0.29$\pm$0.14 & 0.09$\pm$0.06 & 0.07$\pm$0.05 & 0.02$\pm$0.03 & 0.13$\pm$0.12 & 0.15$\pm$0.12 & 0.27$\pm$0.14 & 0.39$\pm$0.16 & 0.37$\pm$0.16 & 0.30$\pm$0.38 \\

 304  & 0.44$\pm$0.13 & 0.55$\pm$0.15 & 0.37$\pm$0.12 & 0.15$\pm$0.08 & 0.12$\pm$0.06 & 0.04$\pm$0.04 & 0.27$\pm$0.13 & 0.26$\pm$0.15 & 0.42$\pm$0.14 & 0.49$\pm$0.16 & 0.50$\pm$0.15 & 0.36$\pm$0.37 \\

 1700  & 0.61$\pm$0.14 & 0.74$\pm$0.15 & 0.67$\pm$0.16 & 0.15$\pm$0.07 & 0.12$\pm$0.06 & 0.07$\pm$0.03 & 0.54$\pm$0.13 & 0.57$\pm$0.13 & 0.50$\pm$0.13 & 0.68$\pm$0.18 & 0.68$\pm$0.17 & 0.45$\pm$0.30 \\

 1600  & 0.63$\pm$0.13 & 0.77$\pm$0.15 & 0.69$\pm$0.16 & 0.15$\pm$0.07 & 0.12$\pm$0.06 & 0.08$\pm$0.04 & 0.54$\pm$0.14 & 0.57$\pm$0.13 & 0.52$\pm$0.13 & 0.71$\pm$0.19 & 0.71$\pm$0.18 & 0.43$\pm$0.29 \\

 1600 1700  & 0.64$\pm$0.13 & 0.78$\pm$0.15 & 0.69$\pm$0.16 & 0.18$\pm$0.07 & 0.15$\pm$0.06 & 0.09$\pm$0.04 & 0.58$\pm$0.13 & 0.61$\pm$0.13 & 0.55$\pm$0.14 & 0.70$\pm$0.19 & 0.70$\pm$0.18 & 0.45$\pm$0.29 \\

 1600 1700 304  & 0.64$\pm$0.14 & 0.79$\pm$0.16 & 0.71$\pm$0.17 & 0.21$\pm$0.08 & 0.19$\pm$0.06 & 0.12$\pm$0.04 & 0.67$\pm$0.13 & 0.68$\pm$0.13 & 0.59$\pm$0.14 & 0.72$\pm$0.20 & 0.73$\pm$0.19 & 0.48$\pm$0.23 \\

 All AIA  & 0.65$\pm$0.13 & 0.80$\pm$0.16 & 0.71$\pm$0.17 & 0.21$\pm$0.08 & 0.18$\pm$0.07 & 0.12$\pm$0.04 & 0.67$\pm$0.13 & 0.68$\pm$0.13 & 0.59$\pm$0.14 & 0.73$\pm$0.19 & 0.73$\pm$0.19 & 0.45$\pm$0.24 \\

 1600 1700 HMI  & 0.65$\pm$0.13 & 0.79$\pm$0.15 & 0.70$\pm$0.16 & 0.21$\pm$0.08 & 0.18$\pm$0.06 & 0.11$\pm$0.04 & 0.64$\pm$0.12 & 0.66$\pm$0.12 & 0.57$\pm$0.13 & 0.72$\pm$0.19 & 0.72$\pm$0.19 & 0.43$\pm$0.26 \\

 1600 1700 304 HMI  & 0.65$\pm$0.13 & 0.80$\pm$0.15 & 0.71$\pm$0.17 & 0.23$\pm$0.08 & 0.20$\pm$0.07 & 0.13$\pm$0.05 & 0.68$\pm$0.13 & 0.69$\pm$0.13 & 0.60$\pm$0.14 & 0.72$\pm$0.19 & 0.73$\pm$0.19 & 0.43$\pm$0.25 \\

 All AIA HMI  & 0.65$\pm$0.12 & 0.80$\pm$0.15 & 0.71$\pm$0.16 & 0.23$\pm$0.08 & 0.20$\pm$0.07 & 0.13$\pm$0.04 & 0.69$\pm$0.13 & 0.69$\pm$0.13 & 0.60$\pm$0.14 & 0.73$\pm$0.20 & 0.73$\pm$0.19 & 0.42$\pm$0.30 \\

 DNN baseline  & 0.49$\pm$0.11 & 0.64$\pm$0.14 & 0.56$\pm$0.15 & 0.14$\pm$0.07 & 0.10$\pm$0.06 & 0.01$\pm$0.02 & 0.29$\pm$0.09 & 0.32$\pm$0.09 & 0.44$\pm$0.13 & 0.55$\pm$0.16 & 0.56$\pm$0.17 & 0.09$\pm$0.06 \\
 
\bottomrule
\end{tabular}
\caption{Mean and standard deviation of the distribution of the Pearson correlation coefficient $<r>$ of the linear correlation  for linear regression between each of the thermodynamic variables obtained by \irissq\ and \atoi\ at optical depths \ltau~=~-5.2,~-4.2,~-3.2 of all the data in the test dataset (144) for each experiment evaluated in this investigation.\label{table:all_lincorr_original}}
\end{sidewaystable}

\begin{sidewaystable}
\begin{tabular}{c *{12}{c}}
\toprule
\multicolumn{1}{c}{Experiment} &
\multicolumn{3}{c}{T} &
\multicolumn{3}{c}{$v_{los}$} &
\multicolumn{3}{c}{$v_{turb}$} &
\multicolumn{3}{c}{$n_{e}$} \\
\cmidrule(lr){2-4} \cmidrule(lr){5-7} \cmidrule(lr){8-10} \cmidrule(lr){11-13}

&
-5.2 & -4.2 & -3.2 &
-5.2 & -4.2 & -3.2 &
-5.2 & -4.2 & -3.2 &
-5.2 & -4.2 & -3.2 \\ 
\cmidrule(lr){2-2} \cmidrule(lr){3-3} \cmidrule(lr){4-4}
\cmidrule(lr){5-5} \cmidrule(lr){6-6} \cmidrule(lr){7-7}
\cmidrule(lr){8-8} \cmidrule(lr){9-9} \cmidrule(lr){10-10}
\cmidrule(lr){11-11}\cmidrule(lr){12-12}\cmidrule(lr){13-13}


 193  & 0.38$\pm$0.16 & 0.42$\pm$0.18 & 0.28$\pm$0.14 & 0.10$\pm$0.08 & 0.07$\pm$0.06 & 0.02$\pm$0.03 & 0.10$\pm$0.10 & 0.12$\pm$0.11 & 0.27$\pm$0.14 & 0.42$\pm$0.17 & 0.40$\pm$0.18 & 0.15$\pm$0.11 \\

 171  & 0.36$\pm$0.15 & 0.38$\pm$0.16 & 0.25$\pm$0.13 & 0.10$\pm$0.07 & 0.07$\pm$0.06 & 0.02$\pm$0.03 & 0.12$\pm$0.12 & 0.14$\pm$0.12 & 0.25$\pm$0.13 & 0.40$\pm$0.15 & 0.35$\pm$0.17 & 0.14$\pm$0.10 \\

 304  & 0.50$\pm$0.15 & 0.54$\pm$0.17 & 0.35$\pm$0.14 & 0.16$\pm$0.09 & 0.11$\pm$0.07 & 0.04$\pm$0.04 & 0.27$\pm$0.13 & 0.26$\pm$0.14 & 0.37$\pm$0.14 & 0.56$\pm$0.15 & 0.52$\pm$0.18 & 0.18$\pm$0.09 \\

 1700  & 0.70$\pm$0.15 & 0.75$\pm$0.17 & 0.70$\pm$0.16 & 0.17$\pm$0.08 & 0.13$\pm$0.07 & 0.07$\pm$0.04 & 0.53$\pm$0.14 & 0.56$\pm$0.13 & 0.51$\pm$0.13 & 0.76$\pm$0.15 & 0.72$\pm$0.17 & 0.32$\pm$0.13 \\

 1600  & 0.72$\pm$0.16 & 0.77$\pm$0.18 & 0.70$\pm$0.17 & 0.16$\pm$0.08 & 0.13$\pm$0.06 & 0.07$\pm$0.04 & 0.53$\pm$0.14 & 0.56$\pm$0.13 & 0.52$\pm$0.13 & 0.78$\pm$0.15 & 0.74$\pm$0.18 & 0.33$\pm$0.09 \\

 1600 1700  & 0.74$\pm$0.16 & 0.78$\pm$0.18 & 0.72$\pm$0.17 & 0.19$\pm$0.08 & 0.15$\pm$0.06 & 0.09$\pm$0.04 & 0.57$\pm$0.13 & 0.60$\pm$0.13 & 0.54$\pm$0.13 & 0.80$\pm$0.15 & 0.76$\pm$0.18 & 0.36$\pm$0.11 \\

 1600 1700 304  & 0.74$\pm$0.16 & 0.80$\pm$0.18 & 0.73$\pm$0.17 & 0.22$\pm$0.09 & 0.19$\pm$0.07 & 0.11$\pm$0.04 & 0.66$\pm$0.13 & 0.67$\pm$0.14 & 0.57$\pm$0.14 & 0.80$\pm$0.15 & 0.78$\pm$0.18 & 0.36$\pm$0.13 \\

 All AIA  & 0.74$\pm$0.16 & 0.80$\pm$0.19 & 0.72$\pm$0.18 & 0.22$\pm$0.09 & 0.19$\pm$0.07 & 0.12$\pm$0.04 & 0.66$\pm$0.13 & 0.67$\pm$0.14 & 0.57$\pm$0.14 & 0.80$\pm$0.16 & 0.78$\pm$0.19 & 0.36$\pm$0.12 \\

 1600 1700 HMI  & 0.74$\pm$0.15 & 0.79$\pm$0.18 & 0.72$\pm$0.17 & 0.22$\pm$0.09 & 0.19$\pm$0.07 & 0.12$\pm$0.04 & 0.63$\pm$0.13 & 0.65$\pm$0.13 & 0.56$\pm$0.12 & 0.81$\pm$0.15 & 0.77$\pm$0.19 & 0.36$\pm$0.10 \\

 1600 1700 304 HMI  & 0.75$\pm$0.15 & 0.80$\pm$0.18 & 0.73$\pm$0.18 & 0.24$\pm$0.10 & 0.21$\pm$0.08 & 0.13$\pm$0.05 & 0.67$\pm$0.13 & 0.68$\pm$0.13 & 0.58$\pm$0.13 & 0.81$\pm$0.14 & 0.78$\pm$0.19 & 0.36$\pm$0.10 \\

 All AIA HMI  & 0.75$\pm$0.15 & 0.80$\pm$0.18 & 0.72$\pm$0.17 & 0.24$\pm$0.10 & 0.20$\pm$0.08 & 0.13$\pm$0.04 & 0.68$\pm$0.13 & 0.69$\pm$0.13 & 0.58$\pm$0.13 & 0.81$\pm$0.14 & 0.78$\pm$0.19 & 0.37$\pm$0.10 \\

 DNN baseline  & 0.57$\pm$0.15 & 0.64$\pm$0.17 & 0.57$\pm$0.16 & 0.14$\pm$0.08 & 0.10$\pm$0.06 & 0.01$\pm$0.02 & 0.30$\pm$0.10 & 0.33$\pm$0.09 & 0.41$\pm$0.14 & 0.62$\pm$0.14 & 0.61$\pm$0.17 & 0.14$\pm$0.05 \\

\bottomrule
\end{tabular}
\caption{Same as Table \ref{table:all_lincorr_original} for the linear correlation between the thermodynamic variables obtained by \irissq\ and \atoi\ after removing the outliers using the Z-score filtering meeting with Z-score=3.\label{table:all_lincorr_Zscore}}
\end{sidewaystable}

\begin{sidewaystable}
\begin{tabular}{c *{12}{c}}
\toprule
\multicolumn{1}{c}{Experiment} &
\multicolumn{3}{c}{T} &
\multicolumn{3}{c}{$v_{los}$} &
\multicolumn{3}{c}{$v_{turb}$} &
\multicolumn{3}{c}{$n_{e}$} \\
\cmidrule(lr){2-4} \cmidrule(lr){5-7} \cmidrule(lr){8-10} \cmidrule(lr){11-13}
&
-5.2 & -4.2 & -3.2 &
-5.2 & -4.2 & -3.2 &
-5.2 & -4.2 & -3.2 &
-5.2 & -4.2 & -3.2 \\ 
\cmidrule(lr){2-2} \cmidrule(lr){3-3} \cmidrule(lr){4-4}
\cmidrule(lr){5-5} \cmidrule(lr){6-6} \cmidrule(lr){7-7}
\cmidrule(lr){8-8} \cmidrule(lr){9-9} \cmidrule(lr){10-10}
\cmidrule(lr){11-11}\cmidrule(lr){12-12}\cmidrule(lr){13-13} 

 193  & 0.38 (49) & 0.42 (39) & 0.28 (54) & 0.10 (81) & 0.07 (91) & 0.02 (100) & 0.10 (65) & 0.12 (65) & 0.27 (53) & 0.42 (42) & 0.40 (44) & 0.15 (64) \\

 171  & 0.36 (47) & 0.38 (40) & 0.25 (48) & 0.10 (84) & 0.07 (94) & 0.02 (100) & 0.12 (65) & 0.14 (64) & 0.25 (55) & 0.40 (48) & 0.35 (41) & 0.14 (78) \\

 304  & 0.50 (44) & 0.54 (48) & 0.35 (46) & 0.16 (69) & 0.11 (85) & 0.04 (99) & 0.27 (58) & 0.26 (49) & 0.37 (46) & 0.56 (53) & 0.52 (45) & 0.18 (84) \\

 1700  & 0.70 (64) & 0.75 (67) & 0.70 (58) & 0.17 (75) & 0.13 (88) & 0.07 (100) & 0.53 (67) & 0.56 (71) & 0.51 (67) & 0.76 (83) & 0.72 (59) & 0.32 (66) \\

 1600  & 0.72 (72) & 0.77 (67) & 0.70 (53) & 0.16 (79) & 0.13 (92) & 0.07 (100) & 0.53 (61) & 0.56 (67) & 0.52 (72) & 0.78 (85) & 0.74 (58) & 0.33 (76) \\

 1600 1700  & 0.74 (75) & 0.78 (64) & 0.72 (55) & 0.19 (78) & 0.15 (88) & 0.09 (100) & 0.57 (69) & 0.60 (74) & 0.54 (72) & 0.80 (88) & 0.76 (59) & 0.36 (69) \\

 1600 1700 304  & 0.74 (72) & 0.80 (65) & 0.73 (54) & 0.22 (74) & 0.19 (84) & 0.11 (100) & 0.66 (85) & 0.67 (84) & 0.57 (65) & 0.80 (88) & 0.78 (54) & 0.36 (69) \\

 All AIA  & 0.74 (73) & 0.80 (63) & 0.72 (52) & 0.22 (71) & 0.19 (81) & 0.12 (99) & 0.66 (81) & 0.67 (82) & 0.57 (62) & 0.80 (87) & 0.78 (54) & 0.36 (69) \\

 1600 1700 HMI  & 0.74 (78) & 0.79 (68) & 0.72 (54) & 0.22 (77) & 0.19 (85) & 0.12 (99) & 0.63 (81) & 0.65 (84) & 0.56 (78) & \textbf{0.81 (88)} & 0.77 (58) & 0.36 (78) \\

 1600 1700 304 HMI  & 0.75 (74) & \textbf{0.80 (68)} & 0.73 (53) & \textbf{0.24 (73)} & 0.21 (80) & 0.13 (95) & 0.67 (88) & 0.68 (87) & 0.58 (69) & \textbf{0.81 (88)} & 0.78 (56) & 0.36 (75) \\

 All AIA HMI  & 0.75 (77) & 0.80 (67) & 0.72 (54) & 0.24 (71) & 0.20 (83) & 0.13 (97) & 0.68 (90) & \textbf{0.69 (90)} & 0.58 (67) & \textbf{0.81 (88)} & 0.78 (57) & 0.37 (74) \\

 DNN baseline  & 0.57 (66) & 0.64 (62) & 0.57 (57) & 0.14 (78) & 0.10 (86) & 0.01 (100) & 0.30 (73) & 0.33 (80) & 0.41 (58) & 0.62 (71) & 0.61 (58) & 0.14 (96) \\

\bottomrule
\end{tabular}
\caption{Mean of $r$ as shown for Table \ref{table:all_lincorr_Zscore}, i.e., for linear correlations after removing the outliers. The number in parenthesis indicates the percentage of the data sets in the test dataset within an interval of $<r>\pm0.1$.\label{table:distribution_mean}}
\end{sidewaystable}

\begin{sidewaystable}
\begin{tabular}{c *{12}{c}}
\toprule
\multicolumn{1}{c}{Experiment} &
\multicolumn{3}{c}{T} &
\multicolumn{3}{c}{$v_{los}$} &
\multicolumn{3}{c}{$v_{turb}$} &
\multicolumn{3}{c}{$n_{e}$} \\
\cmidrule(lr){2-4} \cmidrule(lr){5-7} \cmidrule(lr){8-10} \cmidrule(lr){11-13}
&
-5.2 & -4.2 & -3.2 &
-5.2 & -4.2 & -3.2 &
-5.2 & -4.2 & -3.2 &
-5.2 & -4.2 & -3.2 \\ 
\cmidrule(lr){2-2} \cmidrule(lr){3-3} \cmidrule(lr){4-4}
\cmidrule(lr){5-5} \cmidrule(lr){6-6} \cmidrule(lr){7-7}
\cmidrule(lr){8-8} \cmidrule(lr){9-9} \cmidrule(lr){10-10}
\cmidrule(lr){11-11}\cmidrule(lr){12-12}\cmidrule(lr){13-13}

 193  & 0.45 (44) & 0.42 (39) & 0.28 (54) & 0.06 (75) & 0.04 (85) & 0.01 (100) & 0.06 (63) & 0.07 (60) & 0.20 (44) & 0.48 (46) & 0.37 (39) & 0.10 (58) \\

 171  & 0.26 (40) & 0.30 (37) & 0.17 (47) & 0.02 (63) & 0.01 (77) & 0.00 (99) & 0.13 (65) & 0.14 (65) & 0.15 (39) & 0.28 (37) & 0.24 (34) & 0.07 (57) \\

 304  & 0.43 (44) & 0.53 (47) & 0.25 (40) & 0.12 (65) & 0.10 (83) & 0.01 (97) & 0.23 (52) & 0.29 (49) & 0.46 (47) & 0.45 (38) & 0.45 (38) & 0.17 (84) \\

 1700  & 0.75 (79) & 0.81 (82) & 0.69 (57) & 0.10 (59) & 0.07 (73) & 0.08 (99) & 0.56 (76) & 0.56 (72) & 0.47 (58) & 0.78 (86) & 0.72 (57) & 0.36 (72) \\

 1600  & 0.75 (83) & 0.82 (83) & 0.70 (55) & 0.10 (64) & 0.10 (87) & 0.04 (95) & 0.58 (67) & 0.58 (72) & 0.59 (72) & 0.79 (88) & 0.73 (55) & 0.33 (76) \\

 1600 1700  & 0.77 (85) & 0.82 (85) & 0.71 (53) & 0.18 (77) & 0.14 (88) & 0.09 (100) & 0.58 (68) & 0.59 (70) & 0.57 (80) & 0.80 (88) & 0.83 (81) & 0.36 (69) \\

 1600 1700 304  & 0.77 (85) & 0.82 (84) & 0.72 (53) & 0.18 (67) & 0.16 (83) & 0.10 (100) & 0.63 (69) & 0.63 (63) & 0.59 (71) & 0.81 (88) & 0.84 (82) & 0.36 (68) \\

 All AIA  & 0.77 (85) & 0.83 (84) & 0.80 (77) & 0.17 (63) & 0.17 (83) & 0.09 (96) & 0.63 (69) & 0.63 (65) & 0.51 (45) & 0.81 (88) & 0.83 (82) & 0.35 (67) \\

 1600 1700 HMI  & 0.77 (86) & 0.82 (84) & 0.71 (53) & 0.24 (80) & 0.20 (85) & 0.11 (98) & 0.60 (67) & 0.60 (65) & 0.51 (54) & 0.81 (88) & 0.83 (81) & 0.39 (77) \\

 1600 1700 304 HMI  & 0.78 (85) & \textbf{0.83 (85)} & 0.71 (50) & \textbf{0.27 (72)} & 0.23 (81) & 0.12 (97) & \textbf{0.63 (72)} & 0.63 (65) & 0.52 (51) & 0.81 (88) & \textbf{0.84 (83)} & 0.39 (73) \\

 All AIA HMI  & 0.78 (85) & 0.83 (83) & 0.72 (50) & 0.23 (71) & 0.19 (80) & 0.12 (97) & 0.62 (56) & 0.62 (56) & 0.52 (51) & 0.81 (88) & 0.84 (82) & 0.39 (78) \\

 DNN baseline  & 0.58 (67) & 0.64 (58) & 0.62 (73) & 0.11 (76) & 0.07 (88) & -0.00 (100) & 0.25 (62) & 0.27 (55) & 0.34 (40) & 0.68 (82) & 0.63 (67) & 0.06 (62) \\
 
\bottomrule
\end{tabular}
\caption{Same as Table \ref{table:distribution_mean} for the mode, $m$, instead of the mean, of the distribution of $r$, when the linear correlation ignores the outliers. 
The number in parentheses indicates the percentage of the data sets in the test dataset within an interval of $m\pm0.1$.\label{table:distribution_mode}}
\end{sidewaystable}

\subsection{Contribution of each AIA channel and the HMI magnetogram}

To quantify the contribution of input features to model behavior, we compute feature attributions using Shapley values, a feature attribution method derived from cooperative game theory~\citep{shapley:book1952}. 

The Shapely value of a given feature $\xi_i$, denoted by $\phi_i$, corresponds to the marginal contribution of feature $\xi_i$ to a model evaluation score $r$. This means, that the net score $r$ for the model on dataset $x$ is decomposed into the attribution values as $$r = \phi_0 + \sum_{i} \phi(r,\xi_i),$$ where $\phi_0$ would be the correlation between model output of a baseline input and the target prediction. Hence, each Shapely value $\phi_i$ is the additive contribution of the feature $\xi_i$ to $r$, given $\phi_0$. 

Consider the AIA and HMI image stack $x\in\mathbb{R}^{B\times C_{in}\times H\times W}$, which is the input to the model defined in \S~\ref{sec:model}. The dimension $B$ corresponds to the batch of unique image cubes. We consider the \emph{feature} or the \emph{player} in our Shapely value formalism as all the pixels in a given channel, giving a feature $\xi\in\mathbb{R}^{C_{in}}$, and resulting in $C_{in}$ number of features. For a prediction of $\hat{y}\in\mathbb{R}^{B\times C_{out}\times H\times W}$ and a target of $y\in\mathbb{R}^{B\times C_{out}\times H\times W}$, the Shapely formalism assigns an importance to feature $\xi_i$ given an measure of similarity between $\hat{y}$ and $y$. 

We consider the Pearson correlation between $\hat{y}$ and $y$ ($r$) as our attribution function/model evaluation score. The Shapely formalism defines an attribution ($\phi(r,\xi_i)$) for each feature $\xi_i$ as:
\begin{equation}
    \phi(r,\xi_i) = \sum_{S\subseteq \xi \setminus  {\xi_i}} \frac{|S|!(C_{in}-|S|-1)!}{C_{in}!} \left[r(S\cup{\xi_i}) - r(S)\right],
    \label{eqn:shapley}
\end{equation}
where $\xi$ is the set of all features,  $S$ is a subset of $\xi$ not containing the feature $\xi_i$, $r(S)$ is the Pearson correlation computed with the subset of features $S$, and $|S|$ is the count of number of elements in the set $S$. In general, the attribution function may be able to consider only feature sets, and not vectors. In our formalism, we replace the dropped feature $\xi_i$ with a baseline value corresponding to the mean value of $x$ across all pixels and data samples. The mean represents an uninformed value of the feature $\xi_i$ which is within the sample distribution of $\xi$. The $\xi$ is constructed from $x$ by considering all the pixels across different data samples as a \emph{coalition}, and perturbing the group as a unit to construct the attribution. Computing Shapely values requires evaluating over all the possible subsets of $\xi$. Hence, we compute the attributions using KernelSHAP~\citep{kernelshap} implemented via the Captum interpretability library~\citep{captum}. KernelSHAP solves for a limited Monte Carlo subsets of $\xi$ by posing the problem as a linear regression solution. We refer to \cite{kernelshap} for a detailed explanation of the algorithm.

\begin{figure*}
    \centering
    \includegraphics[width=1\linewidth]{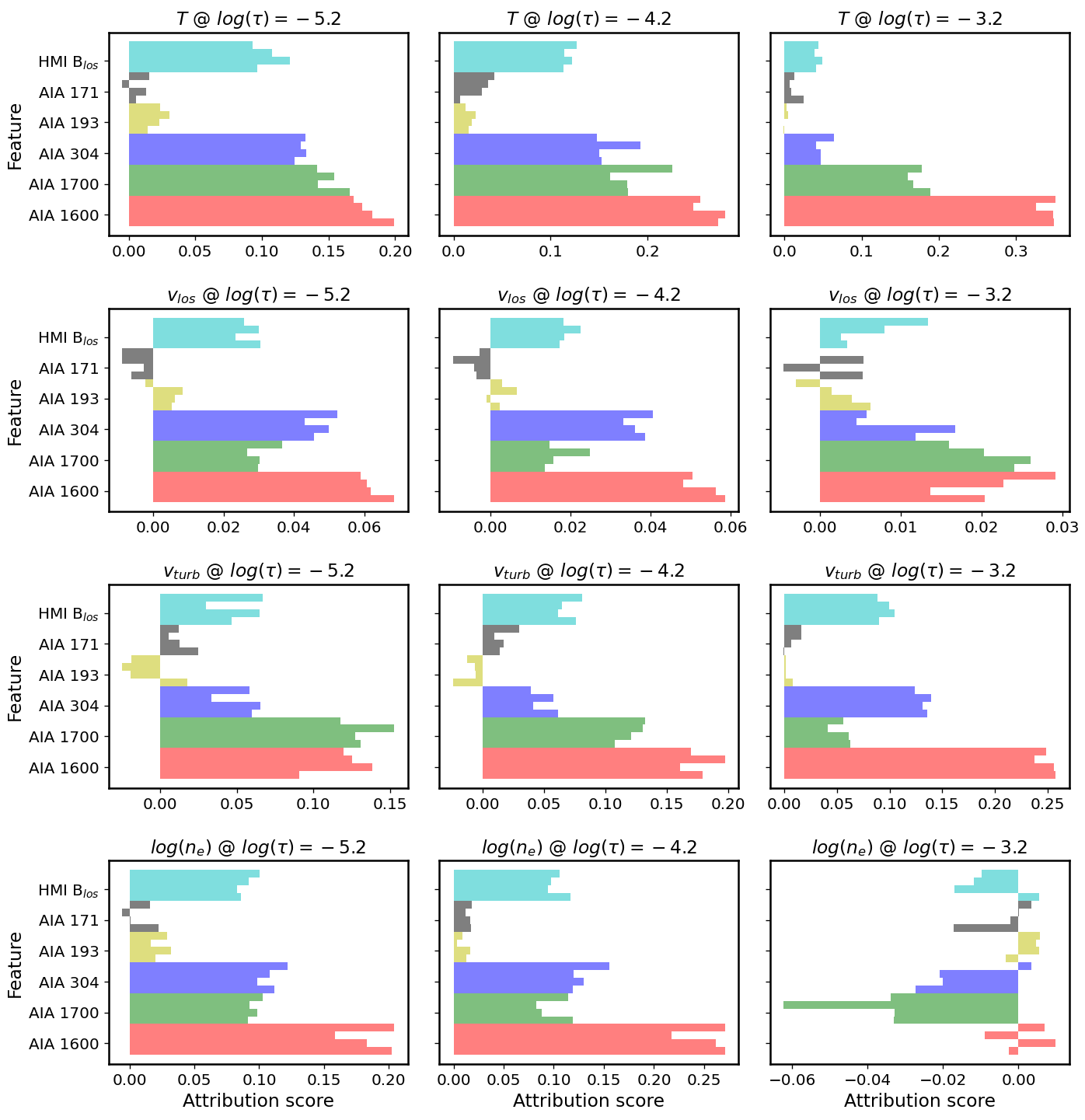}
    \caption{Shapley attribution values for the AIA and HMI passbands for the correlation $r$ between the predicted and target \irissq\ thermodynamic variables.}
    \label{fig:attributions}
\end{figure*}

In our formalism, we have computed the Shapley values for a random set of $100$ data samples of the testing set, and perform the computation $4$ times. This is to ensure robustness in computing the attribution values. The attribution score for each feature of the input for the $4$ sets is shown in Fig.~\ref{fig:attributions}. The different variables are presented with different colors, and the x-axis is the attribution score. This is computed for each of the \irissq\ thermodynamic variables. 


A simple inspection of the panels shown in Figure \ref{fig:attributions} tells us that the most important features contributing to the correlation are the AIA 1600 \AA, AIA 1700 \AA, and the AIA 304 \AA\  channels, followed by the HMI magnetogram. This happens for all thermodynamic variables in all optical depths except for \nne\ at \ltau=-3.2. 

To shed light on these results, we have selected 3 observations that may help us understand the apparently misplaced behavior of \nne at \ltau=-3.2 and other aspects of the results. These observations are shown in the next section, while a more detailed discussion, including all the results shown in this section, is presented in Section \ref{sec:discussion}.

\subsection{Visual comparisons}

\begin{figure*}[h!]
    \centering
    \includegraphics[width=1\linewidth]{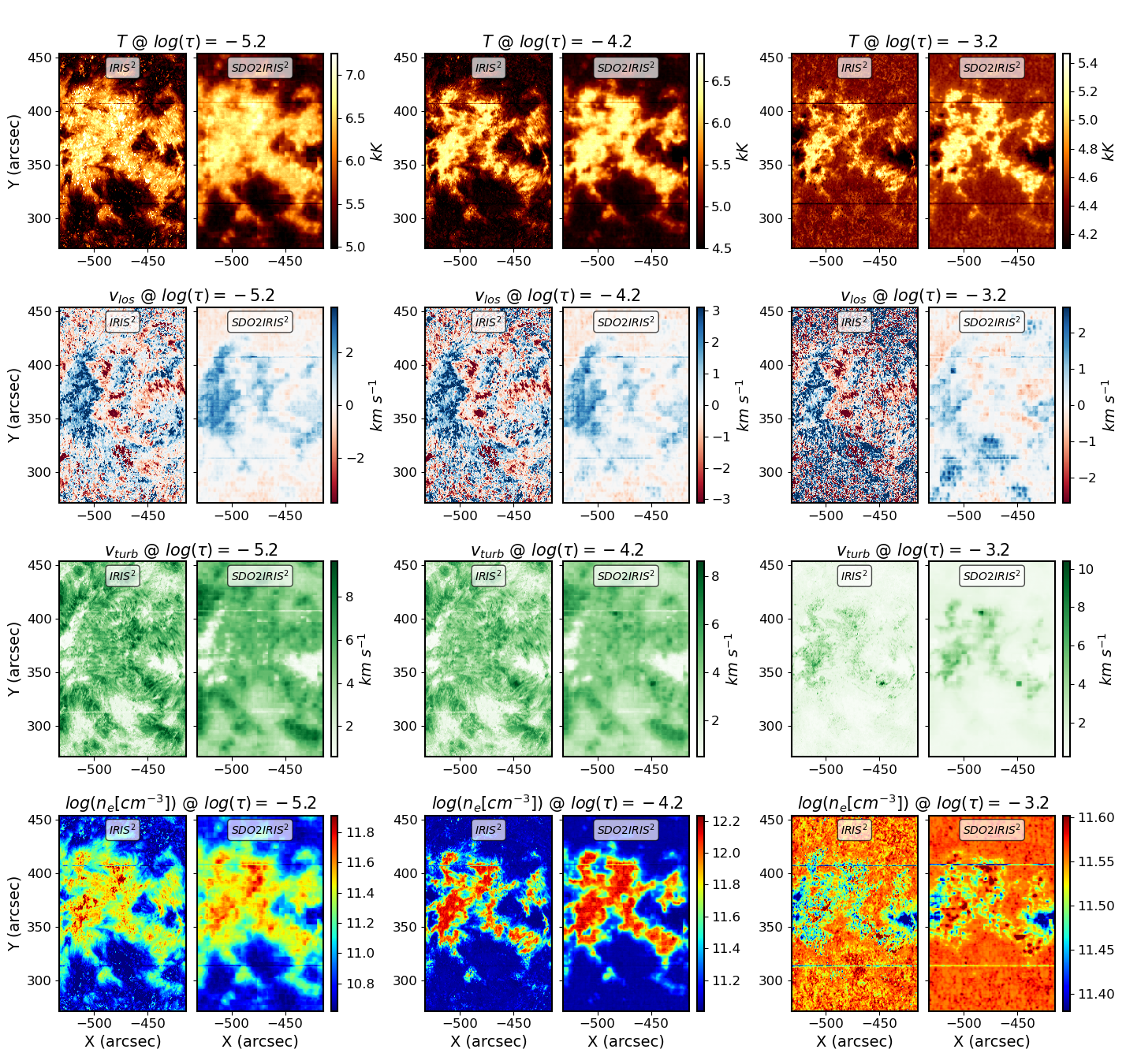}
    \caption{Comparison between the thermodynamic values (in rows) obtained by \irissq\  for the IRIS data observed on 2022-12-29 11:29:33 UT (left panel) and the ones predicted by \atoi\ (right panel) at \ltau=-5.2 (first column), \ltau=-4.2 (second column), and \ltau=-3.2 (third column), by using AIA 1600 \AA, AIA 1700 \AA, AIA 304 \AA, and the HMI magnetogram as the input of \atoi.}
    \label{fig:map1}
\end{figure*}

\begin{figure*}[h!]
    \centering
    \includegraphics[width=1\linewidth]{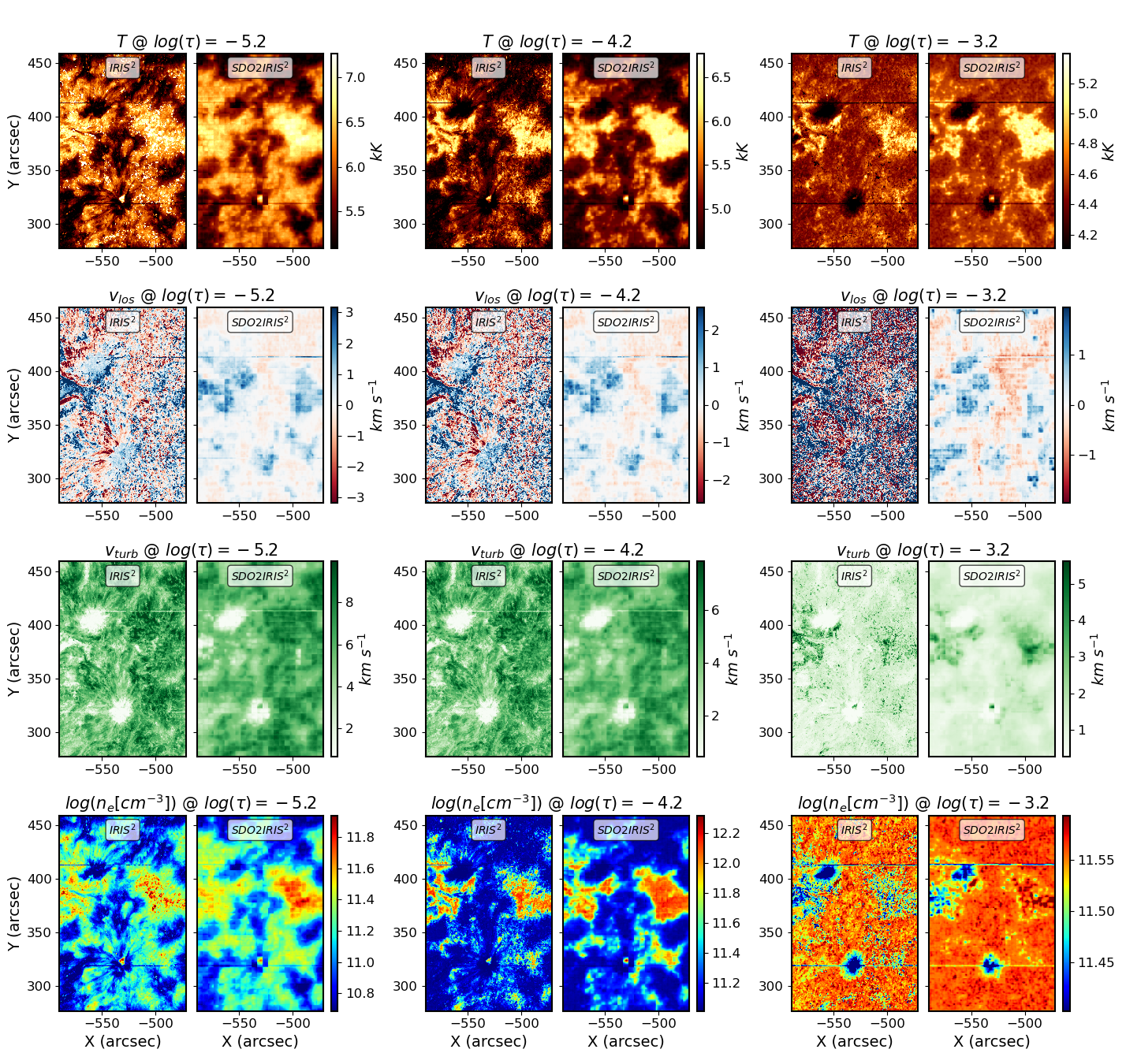}
    \caption{Similar to Figure \ref{fig:map1} for the IRIS data observed on 2023-01-12 11:14:40 UT.}
    \label{fig:map2}
\end{figure*}

\begin{sidewaysfigure*}
    \centering
    \includegraphics[width=\columnwidth]{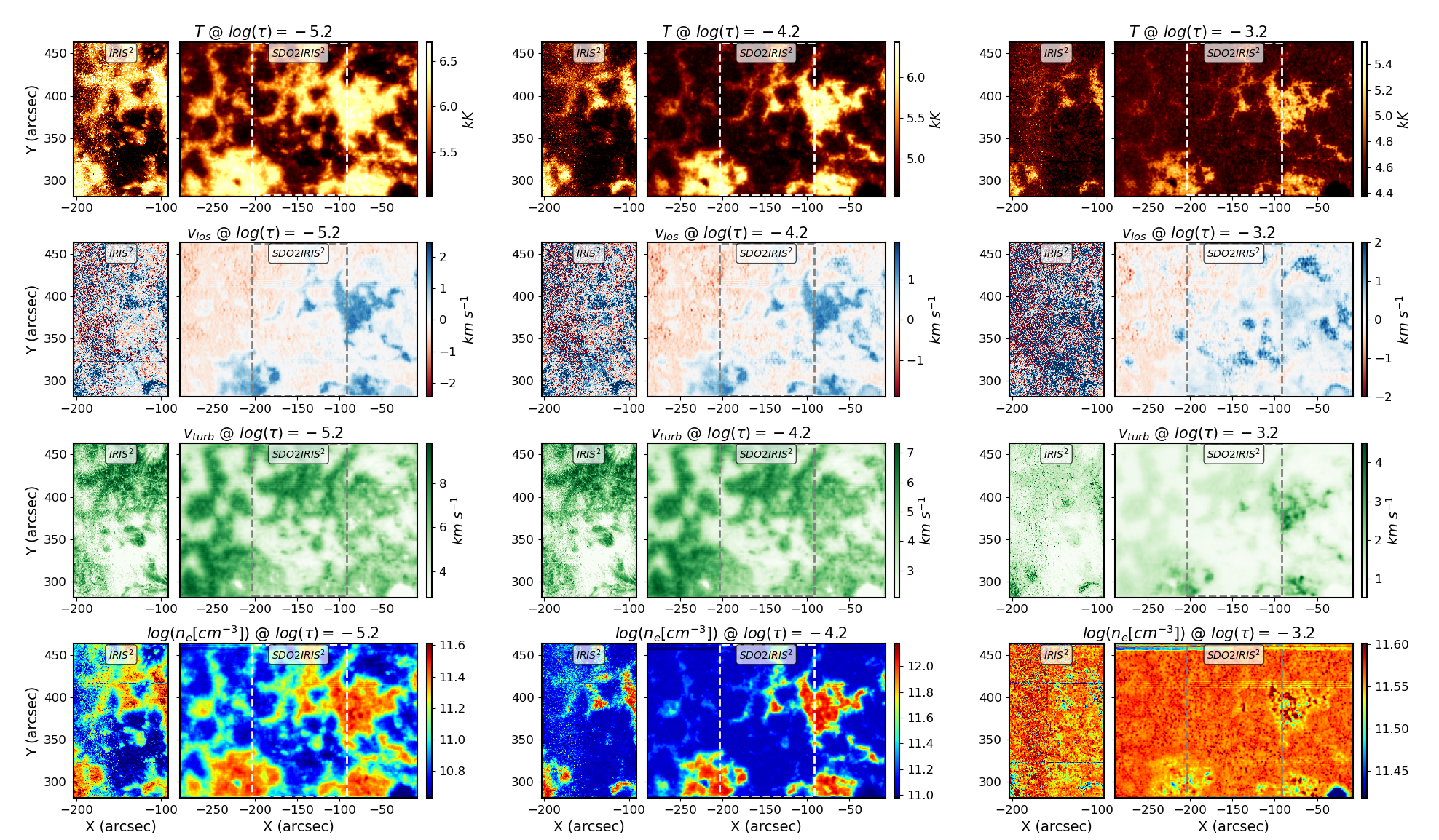}
    \caption{Similar to Figure \ref{fig:map1}, but in this case, the FoV used to predict the thermodynamics from \aia\ and \hmi\ is larger than the one observed by IRIS on 2022-12-08 22:0613. The common FoV observed by both instruments is indicated by a dashed-line rectangle.} \label{fig:map3}
\end{sidewaysfigure*}

Figures \ref{fig:map1}, \ref{fig:map2}, and \ref{fig:map3} show the thermodynamic parameters (rows) obtained by \irissq\ (left sub-panel) and \atoi\ (right sub-panel) in side-by-side panels at the selected optical depths (columns). These figures show some artifacts that \atoi\ may create when predicting targets. The 3 selected datasets are part of the test data set.

Figure \ref{fig:map1} corresponds to the IRIS data observed on 2022-12-08 at 22:06:13 UT and with the IRIS OBSID equal to 3620104077. The field of view (FoV) shows a quiet sun region, plage, one split-umbra sunspot with positive polarity (on the right side of the FoV), and multiple sunspots with negative polarity (on the left side of the FoV) belonging to a complex emerging active region. The T panels show these structures well-defined both in the \irissq\ maps as in the \atoi\ ones at any of the selected optical depths. The light bridge splitting the sunspot located on the right side of the FoV is obvious in the temperature T, especially at \ltau=-5.2 and -4.2; however, in the \atoi\ maps, the light bridge does not completely split the umbra at any optical depth. This discrepancy is likely due to the difference in the spatial sampling between \aia\ and IRIS. The \atoi\ T maps also show hotter predicted values in the plage at all optical depths. 

The \vlos\ maps show a very weak correlation at all optical depths. The main structures seem to keep the same value of \vlos\, but the values obtained by \atoi\ are clearly lower than the \irissq\ ones.

The \atoi\ maps for the \vturb\ miss some structures, e.g., the cluster at (X, Y)=(-460, 390), which can be explained by the spatial resolution. Interestingly, despite the loss of resolution, the values of the \vturb\ obtained by \atoi\ and \irissq\ are rather similar, especially at \ltau=-5.2 and -4.2.

The \nne\ maps at these optical depths are the thermodynamic variables showing the largest $r$ ($\approx0.9$) for this data set, while at \ltau=-3.2, $r\approx0.4$, i.e., a weak correlation. However, the maps show a relatively good correlation in the plage regions and in the sunspots, and a weak correlation in quiet Sun, where the internetwork is barely resolved by \atoi\, while it is resolved by \irissq.

Figure \ref{fig:map2} shows part of the NOAA AR 13180, which was observed by IRIS on 2023-01-02 11:14:40 UT, with IRIS obsid 3610108077. One of the sunspots in this AR, the one located in the lower part of the panels, has a peculiar region in the umbra, at (X, Y) = (-530, 325), that shows an enhancement in T and \vturb. This region actually has \mgii\ profiles compatible with that increase in the T and \vturb. A user not having access to such profiles, and only to the thermodynamics obtained by \atoi\, might consider these values as anomalous - or even wrong - with respect to the general behavior of the neighbour, in this case the umbra. In this case, we suggest to the potential user of \atoi\ to keep a critical view of the results provided by \atoi\, while at the same time considering also other data available by other sources for the FoV studied (such as IRIS$^2$).

Another important artifact that we have to mention is the predicted fiducial lines in the \atoi\ results, especially obvious in the \nne\ at \ltau=-3.2. Since the \irissq\ inverted data used for the training of \atoi\ includes the IRIS fiducial marks, these marks are also {\it trained} and, therefore, predicted by \atoi. These marks may be more obvious in some cases than in others. 

Figure \ref{fig:map3} shows the results of \irissq\ and \atoi\ for the data observed by IRIS on 2022-12-08 at 22:06:13 UT, with IRIS OBSID 3620104077. The general behavior of the thermodynamic variables is similar to that observed in Figure \ref{fig:map1}. We have chosen this dataset because it contains {\it bad pixels} caused by cosmic ray hits on the detectors while IRIS was in the South Atlantic Anomaly (SAA). Examples are seen in a fringe along Y for $-205<X(arcsec)<-150$. The \irissq\ inversions return unrealistic values for those pixels. Nevertheless, the \aia\ data are not affected by the SAA. 

As we mentioned before, we have not included SAA-induced bad pixels in the training dataset for \atoi, since the \aia\ data are not affected by the SAA. Therefore, for the same FoV observed by \aia\ and IRIS, \atoi\ predicts the thermodynamics without the SAA-induced bad values, while \irissq\ does include the effects of the cosmic rays. The right panels in Figure \ref{fig:map1} show the thermodynamic parameters obtained by \atoi\ using as input the from \aia\ and \hmi\ data on 2022-12-08 22:06:13 UT. In this case, we use only one snapshot from these data to compare with the \irissq, while in Figures \ref{fig:map1} and \ref{fig:map2}, as in the training and evaluation of the model's performance, we use the IRIS-slit like AIA data. In this case, we use a single snapshot since that would likely be the way the final user will use \atoi. Moreover, in this example, we have included a different FoV in the input data than the FoV observed by IRIS. Again, this would likely be the way the \atoi\ users will select their own region of interest. By doing this, we can see how \atoi\ will work in a realistic use case. 

In the common area observed by IRIS and \aia\, the thermodynamic values obtained by \irissq\ and the ones predicted by \atoi\ are behaving like in the case described before: a strong correlation for T, with some higher temperatures predicted than observed; weak correlation in the \vlos; rather acceptable values in \vturb; and a strong correlation in \nne. The spatial coherence of the thermodynamic values is very high, which means there is a {\it natural} continuation of the values beyond the common FoV. However, there are some artifacts that need to be mentioned. First, in some cases, there are some horizontal lines showing misplaced values. This is evident for\nne\ at \ltau=-3.2. These lines are spurious predictions of the fiducial lines. As we pointed out before, the fiducial lines may be present in the \atoi\ results, but in this case, because the FoV passed to \atoi\ is larger than the original FoV used during the training of the model, a convolution is needed to build the FoV selected by the \atoi's user. The multiples lines located around the location of the fiducial lines in the data used for training the model are the counterpart of that convolution. 

Another artifact is the presence of misplaced values, or even no values, in the borders of the passed FoV to \atoi. This can be observed in the upper border of the panel showing the \nne\ at \ltau=-3.2. This is due to the extra padding needed to make possible the convolution, taking into account, again, the original size of the \irissq\ inversions to train \atoi. The size of this border effect is determined by the {\tt stride} hyperparameter passed by the user to \atoi: the smaller the stride value is, the smaller the extra padding is. The default value of  {\tt stride} is 4$\times$4. Of course, the solution to this problem is simple: the user interested in a specific FoV can avoid this artifact by selecting a slightly larger FoV, and cropping the \atoi\ results to the FoV of her/his specific interest.


\section{Discussion}\label{sec:discussion}

The results explained above are consistent with the information encoded in the observables and thermodynamic variables used to train \atoi. On one side, we have the \aia\ passband images and the \hmi\ magnetograms (and their locations in the maps), on the other side, the thermodynamic values obtained by \irissq\ (and their locations in the maps). To interpret the results presented above, it is first important to understand where these variables are sensitive to changes in the thermodynamics. 

\begin{table*}[h]
\centering
\begin{tabular}{llll}
\hline
\textbf{Channel} & \textbf{Primary ion(s)} & \textbf{Region of atmosphere*} & \textbf{Char. log$_{10}$(T[K])} \\
\hline
1700\,\AA & continuum & temperature minimum, photosphere & 3.7 \\
304\,\AA & He II & chromosphere, transition region & 4.7 \\
1600\,\AA & C IV + cont. & transition region + upper photosphere & 5.0, 3.7 \\
171\,\AA & Fe IX & quiet corona, upper transition region & 5.8 \\
193\,\AA & Fe XII, XXIV & corona and hot flare plasma & 6.1, 7.3 \\
\hline
\end{tabular}
\caption{Atmospheric channels used in this investigation and corresponding ions, regions, and characteristic temperatures (adapted from Table 1 at \citealt{Lemen12}). (*) Absorption allows imaging of chromospheric material within the corona.}\label{table:aia_response}
\end{table*}

Table \ref{table:aia_response} shows the regions that the \aia\ passbands used to train \atoi\ observe, as well as the primary ion(s), and the ion characteristic temperature. The latter tells us about the average kinetic energy of ions within the plasma, in our case, from the photosphere to the corona. For our investigation, what matters is the region of the atmosphere associated with each channel, and how they contribute to the performance of the \atoi\ model. 

On the other hand, there are the thermodynamic variables obtained from \irissq. For a spectral line, each thermodynamic variable is sensitive to changes at slightly different optical depths. This can be seen in Figure 5 of \citealt{delaCruzRodriguez16}. In that figure, we can also see that the location of these optical depth ranges of sensitivity are slightly different for different solar features. 
Usually, in a FoV, there are several solar features, such as quiet Sun, umbra, penumbra, plage, filament,  etc. Therefore, the uncertainty of the thermodynamic varies spatially on the FoV, and in the optical depth range where it is maximum or minimum. That being said, we can simplify the situation by considering rough optical depths where T, \vlos, \vturb, and \nne\ obtained from the inversion of the \mgii\ lines are  sensitive to changes in the thermodynamics on average. Thus, T is mostly sensitive in all the optical depth ranges selected in this study. It is worth mentioning again that the three optical depths utilized in this work refer 
to a range of $\pm0.2$ around the optical depth shown, e.g., when we talk about ``T at \ltau=-5.2'', it is indeed the average of T in the optical range -5.4 $\leq$ \ltau $\leq$ -5.0. \vlos\ is also sensitive to variations in all the optical depths. However, this variable is the one that has the largest relative uncertainty, i.e., the uncertainty with respect to the value of the \irissq\ variable. 
This is, in part, due to the limited number of representative profiles of the \irissq\ database, and the difficulty of sampling all the velocity space from them. In part, this is likely also related to the fact that Doppler shifts are not necessarily well correlated with AIA images. \vturb\ is sensitive to changes at \ltau=-5.2, in many occasions even more sensitive at \ltau=-4.2, and slightly sensitive at \ltau=-3.2. The \nne\ is mostly sensitive at \ltau=-5.2 and \ltau=-4.2, but barely at \ltau=-3.2. This description refers only to the uncertainty of the thermodynamic variables recovered from the inversion of the \mgii\, and is visually summarized in Figure 4 of \citet{SainzDalda_2023}.

Any combination of \aia\ data that contains encoded physical information (in the integrated intensity within the bandpass) in the solar atmosphere regions that are equivalent to the optical depths where the \irissq\ thermodynamic variables are sensitive to changes in the thermodynamics, will contribute positively to the training of \atoi. The results shown in the previous sections demonstrate that this is the case. Thus, the largest Pearson correlations are reached when the AIA 1600 \AA, AIA 1700 \AA, and AIA 304 \AA\ channels are used together 
to train \atoi, but even when only AIA 1600 \AA\ and AIA 1700 \AA\ are used to train \atoi, some predicted values correlate strongly with the \irissq\ thermodynamic values. The role played by these channels is also clear in the Shapley attribution values shown in Figure \ref{fig:attributions}, where these channels are the largest contributors to the correlation. Interestingly, these channels observe the TR and the upper photosphere and low chromosphere (see Table \ref{table:aia_response}). The \mgii\ lines are not sensitive to the TR, and only the wings of these lines, and the \mguvtt\ lines are able to capture information from the temperature minimum region and the upper photosphere or low chromosphere. The AIA 304 \AA channel is the only one that captures the chromosphere and the TR. While the contribution of this channel to the Pearson correlation seems to be moderate, and slightly increases the correlation of the \atoi\ predictions of AIA 1600 \AA\ and AIA 1700 \AA\ (see Table \ref{table:all_lincorr_Zscore}), the Shapley attribution value of this channel is significant for all the thermodynamic values at any optical depth, being in some case similar or larger than the one for AIA 1700 \AA. If we add the photospheric \hmi\ magnetogram, the correlation improves a little, and especially, some structures are better represented when this information is included. Except for \vlos, the contribution of the \hmi\ magnetogram is similar to that of AIA 304 \AA\ (see Figure \ref{fig:attributions}).

We should mention the Shapley attribution values for \nne\ at \ltau=-3.2. First, the scores are 1 order of magnitude smaller than T and \vturb\ at any optical depth, but similar to ones of \vlos\ at any optical depth. This reflects the weak correlation that exists at these optical depths for \vlos\ and at \ltau=-3.2 for \nne\, between the predicted values by \atoi\ and the ones obtained by \irissq. However, in the case of \vlos, despite the Shapley attribution values being one order of magnitude smaller than for T and \vturb, the behavior of each channel is rather similar, e.g., the largest contributor is AIA 1600 \AA, then AIA 1700 \AA\ (at \ltau=-3.2) and AIA 304 \AA\ (at \ltau=-5.2 and -4.2), and finally the \hmi\ magnetogram.
This means that, despite showing a weak linear correlation, these channels still contribute similarly as they do for T and \vturb, in this case with a weak correlation, but with a similar relative behavior. On the other hand, the Shapley attribution values of these contributors for \nne\ at \ltau=-3.2 show negative values, which means that they contribute to decreasing the correlation of the model to the observations. Our interpretation is that, at this optical depth, \atoi\ is not able to predict \nne\ in the quiet Sun regions (see bottom right sub-panels in Figures \ref{fig:map1}, \ref{fig:map2}, and \ref{fig:map3}), likely because of the lower spatial resolution of AIA. Therefore, these channels are not able to predict well in this region and this decreases the correlation, with AIA 1600 \AA\ and in one case from the \hmi\ magnetogram the only positive contributors. The contribution from AIA 193 \AA\ might look suspicious to the correlation at this optical depth, but it is similar to the values at \ltau=-5.2 and \ltau=-4.2.

\section{Conclusions}\label{sec:conclusions}

We have developed a new tool, named \atoi,  that uses \aia\ images and the \hmi\ magnetograms to predict the thermodynamics in the chromosphere at three optical depths, \ltau = [-5.2, -4.2, -3.2], roughly covering a region from the temperature minimum to the middle and high chromosphere.

We have shown that the predictions of the thermodynamics provided by \atoi\ (using \aia\ 1600 $\AA$, 1700 $\AA$, and 304 $\AA$ bandpass images and \hmi\ magnetograms) when compared with the inversion results obtained by \irissq\ show a strong correlation for T and \nne\ (mode of $r\approx0.80$ in the high-mid and mid chromosphere for $\approx 80\%$ of the data in the test dataset), and a moderate-to-strong correlation in \vturb\ (mode of  $r\approx0.63$ in the high-mid and mid chromosphere for $\approx 70\%$ of the test data). For \vlos\, the correlation is weak, despite being able to capture the general behavior of \vlos\ in some cases, the \vlos\ values predicted by \atoi\ are considerably smaller than the ones obtained by \irissq.

In most cases, including more \aia\ channels than the ones mentioned above may improve the predictions, but only slightly. Nevertheless, while including them is not discouraged, it is not strictly necessary to do so. 

The selected \irissq\ inversions to train \atoi\ do not contain data observed near the limb. Therefore, \atoi\ should not be used to predict the thermodynamics near the limb.

The SDO mission has observed the full-disk Sun since 2010, while the IRIS mission has been operating since 2013 with a smaller field of view (FoV) but higher spatial resolution.  The \atoi\ can be applied to all the data taken by SDO, whether IRIS was observing simultaneously and a larger FoV is required for context, or IRIS was not observing and a proxy of the thermodynamics in the chromosphere is desired. Therefore, \atoi\ adds a new dimension to the results provided by SDO.

The natural evolution of the \atoi\ tool is to consider the inversions obtained with \irissqp\  \citep{SainzDalda_2024, SainzDalda_2026}, i.e., including at least the C II 1334 $\AA$ and 1335 $\AA$ lines, and likely some photospheric lines, to train a new ViT model, AIA2IRIS$^{2+}$. This would allow us to predict the thermodynamics from the mid-to-high photosphere to the high chromosphere.

In summary, \atoi\ provides a rather good estimation of T, \vturb, and \nne\ in the chromosphere, which gives a new usability to the SDO data, and increases the synergy of these data with any other instrument that observes the chromosphere. 

\begin{acknowledgments}

{\it In memoriam:} This article is dedicated to the memory of Alan Title (1938-2026), who, among many other roles and important contributions to solar physics, led the SDO/AIA and IRIS missions. 

The authors gratefully acknowledge support from NASA contract NNG09FA40C (IRIS). IRIS is a NASA small explorer mission developed and operated by LMSAL, with mission operations executed at NASA Ames Research Center, and major contributions to downlink communications funded by ESA and the Norwegian Space Agency. VU would like to acknowledge NASA for support under award number 80NSSC25K7956.
\end{acknowledgments}

\begin{contribution}

ASD is the conceptual author of the \atoi\ tool and led this project. ASD selected the observations and ran the \irissq\ inversions. ASD first introduced the idea of the \atoi\ tool to JK as a project for his high-school internship at LMSAL as part of the {\it Lockheed-Martin - Palo Alto Unified School District internship program}. JK developed, tested, and applied the code to construct IRIS-slit-like AIA rasters. JK ran the first tests using a convolutional neural network (CNN). PSK worked with JK to develop and implement the CNN. VU conceptualized, developed, tested, and evaluated the \atoi\ pipeline, with the core visual transformer (ViT) model. VU investigated other models (CNN, U-Net, and a 1D-DNN as the baseline model) to find the optimal model (ViT) for the project. VU developed the interpretation strategy for the \atoi\ tool. KC collected \hmi\ data corresponding to the IRIS data and co-aligned them. KC also ran the magnetic field extrapolation from the \hmi\ data, which will be used in a spinoff of \atoi. BDP obtained funding for this project, and VH and BDP contributed with helpful ideas and discussions throughout the project. All the co-authors reviewed and contributed to improve the manuscript.


\end{contribution}

%
\facilities{IRIS, SDO/AIA, SDO/HMI, and LMSAL multi-CPU and multi-GPU servers.}

\software{\irissq\ \citep{SainzDalda19}, \href{https://gitlab.com/LMSAL\_HUB/iris\_hub/iris\_lmsalpy}{IRIS-LMSALpy}\footnote{https://gitlab.com/LMSAL\_HUB/iris\_hub/iris\_lmsalpy},  Astropy \citep{2013A&A...558A..33A}, SunPy \citep{sunpy2020}, \citep{Hunter2007}, SciPy \citep{SciPy2020}, Scikit-learn \citep{Pedregosa2011}, OpenCV \citep{Bradski2000}, PyTorch \citep{Paszke2019PyTorch}, Keras \citep{Chollet2015Keras}, Captum~\citep{captum}, Pytorch-lightning~\citep{Falcon_PyTorch_Lightning_2019}, NumPy \citep{Harris20Numpy}, and other basic packages in Python \citep{vanRossum95Python, vanRossum09Python}.}


\bibliography{sample701,python_ML_AI_references,allbib,others_20260127}

@ARTICLE{AsensioRamos08,
       author = {{Asensio Ramos}, A. and {Trujillo Bueno}, J. and {Landi Degl'Innocenti}, E.},
        title = "{Advanced Forward Modeling and Inversion of Stokes Profiles Resulting from the Joint Action of the Hanle and Zeeman Effects}",
      journal = {\apj},
         year = 2008,
        month = aug,
       volume = {683},
       number = {1},
        pages = {542-565},
          doi = {10.1086/589433},
archivePrefix = {arXiv},
       eprint = {0804.2695},
 primaryClass = {astro-ph},
       adsurl = {https://ui.adsabs.harvard.edu/abs/2008ApJ...683..542A}
}

@MISC{AsensioRamos11b,
       author = {{Asensio Ramos}, Andr{\'e}s and {Trujillo Bueno}, Javier and {Landi Degl'Innocenti}, E.},
        title = "{HAZEL: HAnle and ZEeman Light}",
 howpublished = {Astrophysics Source Code Library, record ascl:1109.004},
         year = 2011,
        month = sep,
          eid = {ascl:1109.004},
        pages = {ascl:1109.004},
archivePrefix = {ascl},
       eprint = {1109.004},
       adsurl = {https://ui.adsabs.harvard.edu/abs/2011ascl.soft09004A}
}

@INPROCEEDINGS{Carroll01c,
       author = {{Carroll}, T.~A. and {Balthasar}, H. and {Muglach}, K. and {Nickelt}, I.},
        title = "{Inversion of Stokes Profiles with Artificial Neural Networks}",
    booktitle = {Advanced Solar Polarimetry -- Theory, Observation, and Instrumentation},
         year = 2001,
       editor = {{Sigwarth}, Michael},
       series = {Astronomical Society of the Pacific Conference Series},
       volume = {236},
        month = jan,
        pages = {511},
       adsurl = {https://ui.adsabs.harvard.edu/abs/2001ASPC..236..511C}
}

@ARTICLE{delaCruzRodriguez19,
       author = {{de la Cruz Rodr{\'\i}guez}, J. and {Leenaarts}, J. and {Danilovic}, S. and {Uitenbroek}, H.},
        title = "{STiC: A multiatom non-LTE PRD inversion code for full-Stokes solar observations}",
      journal = {\aap},
         year = 2019,
        month = mar,
       volume = {623},
          eid = {A74},
        pages = {A74},
          doi = {10.1051/0004-6361/201834464},
archivePrefix = {arXiv},
       eprint = {1810.08441},
 primaryClass = {astro-ph.SR},
       adsurl = {https://ui.adsabs.harvard.edu/abs/2019A&A...623A..74D}
}

@ARTICLE{delaCruzRodriguez16,
       author = {{de la Cruz Rodr{\'\i}guez}, Jaime and {Leenaarts}, Jorrit and {Asensio Ramos}, Andr{\'e}s},
        title = "{Non-LTE Inversions of the Mg II h \& k and UV Triplet Lines}",
      journal = {\apjl},
         year = 2016,
        month = oct,
       volume = {830},
       number = {2},
          eid = {L30},
        pages = {L30},
          doi = {10.3847/2041-8205/830/2/L30},
archivePrefix = {arXiv},
       eprint = {1609.09527},
 primaryClass = {astro-ph.SR},
       adsurl = {https://ui.adsabs.harvard.edu/abs/2016ApJ...830L..30D}
}

@ARTICLE{delToroIniesta16,
       author = {{del Toro Iniesta}, Jose Carlos and {Ruiz Cobo}, Basilio},
        title = "{Inversion of the radiative transfer equation for polarized light}",
      journal = {Living Reviews in Solar Physics},
         year = 2016,
        month = nov,
       volume = {13},
       number = {1},
          eid = {4},
        pages = {4},
          doi = {10.1007/s41116-016-0005-2},
archivePrefix = {arXiv},
       eprint = {1610.10039},
 primaryClass = {astro-ph.SR},
       adsurl = {https://ui.adsabs.harvard.edu/abs/2016LRSP...13....4D}
}

@ARTICLE{Hoeksema14,
       author = {{Hoeksema}, J. Todd and {Liu}, Yang and {Hayashi}, Keiji and {Sun}, Xudong and {Schou}, Jesper and {Couvidat}, Sebastien and {Norton}, Aimee and {Bobra}, Monica and {Centeno}, Rebecca and {Leka}, K.~D. and {Barnes}, Graham and {Turmon}, Michael},
        title = "{The Helioseismic and Magnetic Imager (HMI) Vector Magnetic Field Pipeline: Overview and Performance}",
      journal = {\solphys},
         year = 2014,
        month = sep,
       volume = {289},
       number = {9},
        pages = {3483-3530},
          doi = {10.1007/s11207-014-0516-8},
archivePrefix = {arXiv},
       eprint = {1404.1881},
 primaryClass = {astro-ph.SR},
       adsurl = {https://ui.adsabs.harvard.edu/abs/2014SoPh..289.3483H}
}

@ARTICLE{Lagg04,
       author = {{Lagg}, A. and {Woch}, J. and {Krupp}, N. and {Solanki}, S.~K.},
        title = "{Retrieval of the full magnetic vector with the He I multiplet at 1083 nm. Maps of an emerging flux region}",
      journal = {\aap},
         year = 2004,
        month = feb,
       volume = {414},
        pages = {1109-1120},
          doi = {10.1051/0004-6361:20031643},
       adsurl = {https://ui.adsabs.harvard.edu/abs/2004A&A...414.1109L}
}

@ARTICLE{LopezAriste05,
       author = {{L{\'o}pez Ariste}, A. and {Casini}, R.},
        title = "{Inference of the magnetic field in spicules from spectropolarimetry of He I D3}",
      journal = {\aap},
         year = 2005,
        month = jun,
       volume = {436},
       number = {1},
        pages = {325-331},
          doi = {10.1051/0004-6361:20042214},
       adsurl = {https://ui.adsabs.harvard.edu/abs/2005A&A...436..325L}
}

@ARTICLE{RuizCobo22,
       author = {{Ruiz Cobo}, B. and {Quintero Noda}, C. and {Gafeira}, R. and {Uitenbroek}, H. and {Orozco Su{\'a}rez}, D. and {P{\'a}ez Ma{\~n}{\'a}}, E.},
        title = "{DeSIRe: Departure coefficient aided Stokes Inversion based on Response functions}",
      journal = {\aap},
         year = 2022,
        month = apr,
       volume = {660},
          eid = {A37},
        pages = {A37},
          doi = {10.1051/0004-6361/202140877},
archivePrefix = {arXiv},
       eprint = {2202.02226},
 primaryClass = {astro-ph.SR},
       adsurl = {https://ui.adsabs.harvard.edu/abs/2022A&A...660A..37R}
}

@ARTICLE{SainzDalda19,
       author = {{Sainz Dalda}, Alberto and {de la Cruz Rodr{\'\i}guez}, Jaime and {De Pontieu}, Bart and {Go{\v{s}}i{\'c}}, Milan},
        title = "{Recovering Thermodynamics from Spectral Profiles observed by IRIS: A Machine and Deep Learning Approach}",
      journal = {\apjl},
         year = 2019,
        month = apr,
       volume = {875},
       number = {2},
          eid = {L18},
        pages = {L18},
          doi = {10.3847/2041-8213/ab15d9},
archivePrefix = {arXiv},
       eprint = {1904.08390},
 primaryClass = {astro-ph.SR},
       adsurl = {https://ui.adsabs.harvard.edu/abs/2019ApJ...875L..18S}
}

@ARTICLE{Socas-Navarro05a,
       author = {{Socas-Navarro}, H.},
        title = "{Strategies for Spectral Profile Inversion Using Artificial Neural Networks}",
      journal = {\apj},
         year = 2005,
        month = mar,
       volume = {621},
       number = {1},
        pages = {545-553},
          doi = {10.1086/427431},
archivePrefix = {arXiv},
       eprint = {astro-ph/0410567},
 primaryClass = {astro-ph},
       adsurl = {https://ui.adsabs.harvard.edu/abs/2005ApJ...621..545S}
}

@ARTICLE{Socas-Navarro15,
       author = {{Socas-Navarro}, H. and {de la Cruz Rodr{\'\i}guez}, J. and {Asensio Ramos}, A. and {Trujillo Bueno}, J. and {Ruiz Cobo}, B.},
        title = "{An open-source, massively parallel code for non-LTE synthesis and inversion of spectral lines and Zeeman-induced Stokes profiles}",
      journal = {\aap},
         year = 2015,
        month = may,
       volume = {577},
          eid = {A7},
        pages = {A7},
          doi = {10.1051/0004-6361/201424860},
archivePrefix = {arXiv},
       eprint = {1408.6101},
 primaryClass = {astro-ph.SR},
       adsurl = {https://ui.adsabs.harvard.edu/abs/2015A&A...577A...7S}
}

@ARTICLE{DePontieu14a,
       author = {{De Pontieu}, B. and {Title}, A.~M. and {Lemen}, J.~R. and {Kushner}, G.~D. and {Akin}, D.~J. and {Allard}, B. and {Berger}, T. and {Boerner}, P. and {Cheung}, M. and {Chou}, C. and {Drake}, J.~F. and {Duncan}, D.~W. and {Freeland}, S. and {Heyman}, G.~F. and {Hoffman}, C. and {Hurlburt}, N.~E. and {Lindgren}, R.~W. and {Mathur}, D. and {Rehse}, R. and {Sabolish}, D. and {Seguin}, R. and {Schrijver}, C.~J. and {Tarbell}, T.~D. and {W{\"u}lser}, J. -P. and {Wolfson}, C.~J. and {Yanari}, C. and {Mudge}, J. and {Nguyen-Phuc}, N. and {Timmons}, R. and {van Bezooijen}, R. and {Weingrod}, I. and {Brookner}, R. and {Butcher}, G. and {Dougherty}, B. and {Eder}, J. and {Knagenhjelm}, V. and {Larsen}, S. and {Mansir}, D. and {Phan}, L. and {Boyle}, P. and {Cheimets}, P.~N. and {DeLuca}, E.~E. and {Golub}, L. and {Gates}, R. and {Hertz}, E. and {McKillop}, S. and {Park}, S. and {Perry}, T. and {Podgorski}, W.~A. and {Reeves}, K. and {Saar}, S. and {Testa}, P. and {Tian}, H. and {Weber}, M. and {Dunn}, C. and {Eccles}, S. and {Jaeggli}, S.~A. and {Kankelborg}, C.~C. and {Mashburn}, K. and {Pust}, N. and {Springer}, L. and {Carvalho}, R. and {Kleint}, L. and {Marmie}, J. and {Mazmanian}, E. and {Pereira}, T.~M.~D. and {Sawyer}, S. and {Strong}, J. and {Worden}, S.~P. and {Carlsson}, M. and {Hansteen}, V.~H. and {Leenaarts}, J. and {Wiesmann}, M. and {Aloise}, J. and {Chu}, K. -C. and {Bush}, R.~I. and {Scherrer}, P.~H. and {Brekke}, P. and {Martinez-Sykora}, J. and {Lites}, B.~W. and {McIntosh}, S.~W. and {Uitenbroek}, H. and {Okamoto}, T.~J. and {Gummin}, M.~A. and {Auker}, G. and {Jerram}, P. and {Pool}, P. and {Waltham}, N.},
        title = "{The Interface Region Imaging Spectrograph (IRIS)}",
      journal = {\solphys},
         year = 2014,
        month = jul,
       volume = {289},
       number = {7},
        pages = {2733-2779},
          doi = {10.1007/s11207-014-0485-y},
archivePrefix = {arXiv},
       eprint = {1401.2491},
 primaryClass = {astro-ph.SR},
       adsurl = {https://ui.adsabs.harvard.edu/abs/2014SoPh..289.2733D}
}

@ARTICLE{Lemen12,
       author = {{Lemen}, James R. and {Title}, Alan M. and {Akin}, David J. and {Boerner}, Paul F. and {Chou}, Catherine and {Drake}, Jerry F. and {Duncan}, Dexter W. and {Edwards}, Christopher G. and {Friedlaender}, Frank M. and {Heyman}, Gary F. and {Hurlburt}, Neal E. and {Katz}, Noah L. and {Kushner}, Gary D. and {Levay}, Michael and {Lindgren}, Russell W. and {Mathur}, Dnyanesh P. and {McFeaters}, Edward L. and {Mitchell}, Sarah and {Rehse}, Roger A. and {Schrijver}, Carolus J. and {Springer}, Larry A. and {Stern}, Robert A. and {Tarbell}, Theodore D. and {Wuelser}, Jean-Pierre and {Wolfson}, C. Jacob and {Yanari}, Carl and {Bookbinder}, Jay A. and {Cheimets}, Peter N. and {Caldwell}, David and {Deluca}, Edward E. and {Gates}, Richard and {Golub}, Leon and {Park}, Sang and {Podgorski}, William A. and {Bush}, Rock I. and {Scherrer}, Philip H. and {Gummin}, Mark A. and {Smith}, Peter and {Auker}, Gary and {Jerram}, Paul and {Pool}, Peter and {Soufli}, Regina and {Windt}, David L. and {Beardsley}, Sarah and {Clapp}, Matthew and {Lang}, James and {Waltham}, Nicholas},
        title = "{The Atmospheric Imaging Assembly (AIA) on the Solar Dynamics Observatory (SDO)}",
      journal = {\solphys},
         year = 2012,
        month = jan,
       volume = {275},
       number = {1-2},
        pages = {17-40},
          doi = {10.1007/s11207-011-9776-8},
       adsurl = {https://ui.adsabs.harvard.edu/abs/2012SoPh..275...17L}
}

@ARTICLE{Pesnell12,
       author = {{Pesnell}, W. Dean and {Thompson}, B.~J. and {Chamberlin}, P.~C.},
        title = "{The Solar Dynamics Observatory (SDO)}",
      journal = {\solphys},
         year = 2012,
        month = jan,
       volume = {275},
       number = {1-2},
        pages = {3-15},
          doi = {10.1007/s11207-011-9841-3},
       adsurl = {https://ui.adsabs.harvard.edu/abs/2012SoPh..275....3P}
}

@article{SainzDalda_2026,
doi = {10.3847/1538-4357/ae274c},
url = {https://doi.org/10.3847/1538-4357/ae274c},
year = {2026},
month = {jan},
publisher = {The American Astronomical Society},
volume = {997},
number = {2},
pages = {229},
author = {Sainz Dalda, Alberto and de la Cruz Rodríguez, Jaime and Hansteen, Viggo and De Pontieu, Bart and Gošić, Milan},
title = {The IRIS2+ Inversion Tool: Recovering the Radiative Losses and the Thermodynamics in the Lower Solar Atmosphere},
journal = {The Astrophysical Journal}
}

@ARTICLE{SainzDalda_2024,
       author = {{Sainz Dalda}, Alberto and {Agrawal}, Aaryan and {De Pontieu}, Bart and {Go{\v{s}}i{\'c}}, Milan},
        title = "{IRIS$^{2+}$: A Comprehensive Database of Stratified Thermodynamic Models in the Low Solar Atmosphere}",
      journal = {\apjs},
         year = 2024,
        month = mar,
       volume = {271},
       number = {1},
          eid = {24},
        pages = {24},
          doi = {10.3847/1538-4365/ad1e55},
archivePrefix = {arXiv},
       eprint = {2211.09103},
 primaryClass = {astro-ph.SR},
       adsurl = {https://ui.adsabs.harvard.edu/abs/2024ApJS..271...24S}
}

@ARTICLE{Milic_2018,
       author = {{Mili{\'c}}, I. and {van Noort}, M.},
        title = "{Spectropolarimetric NLTE inversion code SNAPI}",
      journal = {\aap},
         year = 2018,
        month = sep,
       volume = {617},
          eid = {A24},
        pages = {A24},
          doi = {10.1051/0004-6361/201833382},
archivePrefix = {arXiv},
       eprint = {1806.08134},
 primaryClass = {astro-ph.SR},
       adsurl = {https://ui.adsabs.harvard.edu/abs/2018A&A...617A..24M}
}

@ARTICLE{Li_2022,
       author = {{Li}, H. and {del Pino Alem{\'a}n}, T. and {Trujillo Bueno}, J. and {Casini}, R.},
        title = "{TIC: A Stokes Inversion Code for Scattering Polarization with Partial Frequency Redistribution and Arbitrary Magnetic Fields}",
      journal = {\apj},
         year = 2022,
        month = jul,
       volume = {933},
       number = {2},
          eid = {145},
        pages = {145},
          doi = {10.3847/1538-4357/ac745c},
archivePrefix = {arXiv},
       eprint = {2205.15666},
 primaryClass = {astro-ph.SR},
       adsurl = {https://ui.adsabs.harvard.edu/abs/2022ApJ...933..145L}
}

@ARTICLE{SainzDalda_2023,
       author = {{Sainz Dalda}, Alberto and {De Pontieu}, Bart},
        title = "{Recovering Thermodynamics from Spectral Profiles Observed by IRIS. (II). Improved Calculation of the Uncertainties Based on Monte Carlo Experiments}",
      journal = {\apj},
         year = 2023,
        month = feb,
       volume = {944},
       number = {2},
          eid = {118},
        pages = {118},
          doi = {10.3847/1538-4357/acb2c7},
archivePrefix = {arXiv},
       eprint = {2211.01563},
 primaryClass = {astro-ph.SR},
       adsurl = {https://ui.adsabs.harvard.edu/abs/2023ApJ...944..118S}
}

@ARTICLE{Jejcic2022,
  author = {{Jej{\v{c}}i{\v{c}}}, S. and {Heinzel}, P. and {Schmieder}, B. and {Gun{\'a}r}, S. and {Mein}, P. and {Mein}, N. and {Ruan}, G.},
  title = "{Non-LTE Inversion of Prominence Spectroscopic Observations in H$\alpha$ and Mg II h{\&}k lines}",
  journal = {\apj},
  year = 2022,
  volume = 932,
  number = 1,
  pages = {3},
  doi = {10.3847/1538-4357/ac6bf5}
}

@ARTICLE{Molowny-Horas99,
       author = {{Molowny-Horas}, R. and {Heinzel}, P. and {Mein}, P. and {Mein}, N.},
        title = "{A non-LTE inversion procedure for chromospheric cloud-like features}",
      journal = {\aap},
         year = 1999,
        month = may,
       volume = {345},
        pages = {618-628},
       adsurl = {https://ui.adsabs.harvard.edu/abs/1999A&A...345..618M}
}

@ARTICLE{Tziotziou01,
       author = {{Tziotziou}, K. and {Heinzel}, P. and {Mein}, P. and {Mein}, N.},
        title = "{Non-LTE inversion of chromospheric \{\textbackslashCa Ii\} cloud-like features}",
      journal = {\aap},
         year = 2001,
        month = feb,
       volume = {366},
        pages = {686-698},
          doi = {10.1051/0004-6361:20000257},
       adsurl = {https://ui.adsabs.harvard.edu/abs/2001A&A...366..686T}
}

@ARTICLE{sunpy2020,
  author       = {{SunPy Community} and Barnes, Will T. and Bobra, Monica G. and Christe, Steven D. and Freij, Nabil and Hayes, Laura A. and Ireland, Jack and Mumford, Stuart and P{\'e}rez-Su{\'a}rez, David and Ryan, Daniel F. and Shih, Albert Y. and Chanda, Prateek and Glogowski, Kolja and Hewett, Russell and Hughitt, V. Keith and Hill, Andrew and Hiware, Kaustubh and Inglis, Andrew and Kirk, Michael S. F. and Konge, Sudarshan and Mason, James Paul and Maloney, Shane Anthony and Murray, Sophie A. and Panda, Asish and Park, Jongyeob and Pereira, Tiago M. D. and Reardon, Kevin and Savage, Sabrina and Sip{\H o}cz, Brigitta M. and Stansby, David and Jain, Yash and Taylor, Garrison and Yadav, Tannmay and Rajul and Dang, Trung Kien},
  title        = {The SunPy Project: Open Source Development and Status of the Version 1.0 Core Package},
  journal      = {The Astrophysical Journal},
  year         = {2020},
  month        = feb,
  volume       = {890},
  number       = {1},
  pages        = {68},
  doi          = {10.3847/1538-4357/ab4f7a},
  url          = {https://doi.org/10.3847/1538-4357/ab4f7a}
}

@ARTICLE{Hunter2007,
  author       = {Hunter, John D.},
  title        = {Matplotlib: A 2D Graphics Environment},
  journal      = {Computing in Science \& Engineering},
  year         = {2007},
  volume       = {9},
  number       = {3},
  pages        = {90--95}
}

@ARTICLE{Harris20Numpy, 
  author  = {Harris, Charles R. and Millman, K. Jarrod and van der Walt, Stéfan J and Gommers, Ralf and Virtanen, Pauli and Cournapeau, David and Wieser, Eric and Taylor, Julian and Berg, Sebastian and Smith, Nathaniel J. and Kern, Robert and Picus, Matti and Hoyer, Stephan and van Kerkwijk, Marten H. and Brett, Matthew and Haldane, Allan and Fernández del Río, Jaime and Wiebe, Mark and Peterson, Pearu and Gérard-Marchant, Pierre and Sheppard, Kevin and Reddy, Tyler and Weckesser, Warren and Abbasi, Hameer and Gohlke, Christoph and Oliphant, Travis E.},
  title   = {Array programming with {NumPy}}, 
  journal = {Nature}, 
  year    = {2020}, 
  volume  = {585}, 
  pages   = {357–362}, 
  doi     = {10.1038/s41586-020-2649-2} 
}

@book{vanRossum95Python, 
  title={Python reference manual}, 
  author={Van Rossum, Guido and Drake Jr, Fred L}, 
  year={1995}, 
  publisher={Centrum voor Wiskunde en Informatica Amsterdam} 
}

@book{vanRossum09Python, 
 author = {Van Rossum, Guido and Drake, Fred L.}, 
 title = {Python 3 Reference Manual}, 
 year = {2009}, 
 isbn = {1441412697}, 
 publisher = {CreateSpace}, 
 address = {Scotts Valley, CA} 
}

@ARTICLE{SciPy2020,
  author       = {Virtanen, Pauli and Gommers, Ralf and Oliphant, Travis E. and Haberland, Matt and Reddy, Tyler and Cournapeau, David and Burovski, Evgeni and Peterson, Pearu and Weckesser, Warren and Bright, Jonathan and {van der Walt}, St{\'e}fan J. and Brett, Matthew and Wilson, Joshua and Millman, K. Jarrod and Mayorov, Nikolay and Nelson, Andrew R. J. and others},
  title        = {SciPy 1.0: Fundamental Algorithms for Scientific Computing in Python},
  journal      = {Nature Methods},
  year         = {2020},
  volume       = {17},
  number       = {3},
  pages        = {261--272},
  doi          = {10.1038/s41592-019-0686-2},
  url          = {https://doi.org/10.1038/s41592-019-0686-2}
}

@ARTICLE{Pedregosa2011,
  author       = {Pedregosa, Fabian and Varoquaux, Ga{\"e}l and Gramfort, Alexandre and Michel, Vincent and Thirion, Bertrand and Grisel, Olivier and Blondel, Mathieu and Prettenhofer, Peter and Weiss, Ron and Dubourg, Vincent and VanderPlas, Jake and Passos, Alexandre and Cournapeau, David and Brucher, Matthieu and Perrot, Matthieu and Duchesnay, {\'E}douard},
  title        = {Scikit-learn: Machine Learning in Python},
  journal      = {Journal of Machine Learning Research},
  year         = {2011},
  volume       = {12},
  number       = {85},
  pages        = {2825--2830},
  url          = {https://jmlr.org/papers/v12/pedregosa11a.html}
}

@ARTICLE{Bradski2000,
  author       = {Bradski, Gary},
  title        = {The OpenCV Library},
  journal      = {Dr. Dobb's Journal of Software Tools},
  year         = {2000}
}

@MISC{Chollet2015Keras,
  author       = {Chollet, Fran\c{c}ois and others},
  title        = {Keras},
  year         = {2015},
  howpublished = {\url{https://github.com/keras-team/keras}},
  note         = {Accessed 2026-02-24}
}

@INPROCEEDINGS{Paszke2019PyTorch,
  author       = {Paszke, Adam and Gross, Sam and Massa, Francisco and Lerer, Adam and Bradbury, James and Chanan, Gregory and Killeen, Trevor and Lin, Zeming and Gimelshein, Natalia and Antiga, Luca and Desmaison, Alban and K{\"o}pf, Andreas and Yang, Edward and DeVito, Zach and Raison, Martin and Tejani, Alykhan and Chilamkurthy, Sasank and Steiner, Benoit and Fang, Lu and Bai, Junjie and Chintala, Soumith},
  title        = {PyTorch: An Imperative Style, High-Performance Deep Learning Library},
  booktitle    = {Advances in Neural Information Processing Systems 32 (NeurIPS 2019)},
  year         = {2019},
  url          = {https://arxiv.org/abs/1912.01703}
}

@ARTICLE{captum,
       author = {{Kokhlikyan}, Narine and {Miglani}, Vivek and {Martin}, Miguel and {Wang}, Edward and {Alsallakh}, Bilal and {Reynolds}, Jonathan and {Melnikov}, Alexander and {Kliushkina}, Natalia and {Araya}, Carlos and {Yan}, Siqi and {Reblitz-Richardson}, Orion},
        title = "{Captum: A unified and generic model interpretability library for PyTorch}",
      journal = {arXiv e-prints},
         year = 2020,
        month = sep,
          eid = {arXiv:2009.07896},
        pages = {arXiv:2009.07896},
          doi = {10.48550/arXiv.2009.07896},
archivePrefix = {arXiv},
       eprint = {2009.07896},
 primaryClass = {cs.LG},
       adsurl = {https://ui.adsabs.harvard.edu/abs/2020arXiv200907896K}
}

@software{Falcon_PyTorch_Lightning_2019,
author = {Falcon, William and {The PyTorch Lightning team}},
doi = {10.5281/zenodo.3828935},
license = {Apache-2.0},
month = mar,
title = {{PyTorch Lightning}},
url = {https://github.com/Lightning-AI/lightning},
version = {1.4},
year = {2019}
}

@article{Roy2026,
  author  = {Roy, Sujit and Hegde, Dinesha V. and Schmude, Johannes and Lal, Rohit and Gaur, Vishal and Lin, Amy and Mandal, Kshitiz and Singh, Talwinder and Mu{\~n}oz-Jaramillo, Andr{\'e}s and Yang, Kang and Pandey, Chetraj and Hong, Jinsu and Aydin, Berkay and McGranaghan, Ryan and Kasapis, Spiridon and Upendran, Vishal and Bahauddin, Shah and da Silva, Daniel and Freitag, Marcus and Gurung, Iksha and Pogorelov, Nikolai and Watson, Campbell and Maskey, Manil and Bernabe-Moreno, Juan and Ramachandran, Rahul},
  title   = {SuryaBench: Benchmark Dataset for Advancing Machine Learning in Heliophysics and Space Weather Prediction},
  journal = {Scientific Data},
  year    = {2026},
  issn    = {2052-4463},
  doi     = {10.1038/s41597-026-06552-5},
  url     = {https://doi.org/10.1038/s41597-026-06552-5},
}

@ARTICLE{Roy_FM,
       author = {{Roy}, Sujit and {Schmude}, Johannes and {Lal}, Rohit and {Gaur}, Vishal and {Freitag}, Marcus and {Kuehnert}, Julian and {van Kessel}, Theodore and {Hegde}, Dinesha V. and {Mu{\~n}oz-Jaramillo}, Andr{\'e}s and {Jakubik}, Johannes and {Vos}, Etienne and {Mandal}, Kshitiz and {Akbari Asanjan}, Ata and {de Sousa Almeida}, Joao Lucas and {Lin}, Amy and {Singh}, Talwinder and {Yang}, Kang and {Pandey}, Chetraj and {Hong}, Jinsu and {Aydin}, Berkay and {Kurth}, Thorsten and {McGranaghan}, Ryan and {Kasapis}, Spiridon and {Upendran}, Vishal and {Bahauddin}, Shah and {da Silva}, Daniel and {Pogorelov}, Nikolai V. and {Spalding}, Anne and {Watson}, Campbell and {Maskey}, Manil and {Guhathakurta}, Madhulika and {Bernabe-Moreno}, Juan and {Ramachandran}, Rahul},
        title = "{Surya: Foundation Model for Heliophysics}",
      journal = {arXiv e-prints},
         year = 2025,
        month = aug,
          eid = {arXiv:2508.14112},
        pages = {arXiv:2508.14112},
          doi = {10.48550/arXiv.2508.14112},
archivePrefix = {arXiv},
       eprint = {2508.14112},
 primaryClass = {astro-ph.SR},
       adsurl = {https://ui.adsabs.harvard.edu/abs/2025arXiv250814112R}
}

@ARTICLE{kernelshap,
       author = {{Lundberg}, Scott and {Lee}, Su-In},
        title = "{A Unified Approach to Interpreting Model Predictions}",
      journal = {arXiv e-prints},
         year = 2017,
        month = may,
          eid = {arXiv:1705.07874},
        pages = {arXiv:1705.07874},
          doi = {10.48550/arXiv.1705.07874},
archivePrefix = {arXiv},
       eprint = {1705.07874},
 primaryClass = {cs.AI},
       adsurl = {https://ui.adsabs.harvard.edu/abs/2017arXiv170507874L}
}

@incollection{shapley:book1952,
  title = {A Value for n-Person Games},
  author = {Shapley, Lloyd S},
  booktitle = {Contributions to the Theory of Games II},
  editor = {Kuhn, Harold W. and Tucker, Albert W.},
  pages = {307--317},
  year = {1953},
  publisher = {Princeton University Press},
  address = {Princeton}
}

@ARTICLE{layernorm,
       author = {{Xiong}, Ruibin and {Yang}, Yunchang and {He}, Di and {Zheng}, Kai and {Zheng}, Shuxin and {Xing}, Chen and {Zhang}, Huishuai and {Lan}, Yanyan and {Wang}, Liwei and {Liu}, Tie-Yan},
        title = "{On Layer Normalization in the Transformer Architecture}",
      journal = {arXiv e-prints},
         year = 2020,
        month = feb,
          eid = {arXiv:2002.04745},
        pages = {arXiv:2002.04745},
          doi = {10.48550/arXiv.2002.04745},
archivePrefix = {arXiv},
       eprint = {2002.04745},
 primaryClass = {cs.LG},
       adsurl = {https://ui.adsabs.harvard.edu/abs/2020arXiv200204745X}
}

@article{kingma2014adam,
  title={Adam: A method for stochastic optimization},
  author={Kingma, Diederik P and Ba, Jimmy},
  journal={arXiv preprint arXiv:1412.6980},
  year={2014}
}

@ARTICLE{posenc_2017,
       author = {{Gehring}, Jonas and {Auli}, Michael and {Grangier}, David and {Yarats}, Denis and {Dauphin}, Yann N.},
        title = "{Convolutional Sequence to Sequence Learning}",
      journal = {arXiv e-prints},
         year = 2017,
        month = may,
          eid = {arXiv:1705.03122},
        pages = {arXiv:1705.03122},
          doi = {10.48550/arXiv.1705.03122},
archivePrefix = {arXiv},
       eprint = {1705.03122},
 primaryClass = {cs.CL},
       adsurl = {https://ui.adsabs.harvard.edu/abs/2017arXiv170503122G}
}

@ARTICLE{vit_2020,
       author = {{Dosovitskiy}, Alexey and {Beyer}, Lucas and {Kolesnikov}, Alexander and {Weissenborn}, Dirk and {Zhai}, Xiaohua and {Unterthiner}, Thomas and {Dehghani}, Mostafa and {Minderer}, Matthias and {Heigold}, Georg and {Gelly}, Sylvain and {Uszkoreit}, Jakob and {Houlsby}, Neil},
        title = "{An Image is Worth 16x16 Words: Transformers for Image Recognition at Scale}",
      journal = {arXiv e-prints},
         year = 2020,
        month = oct,
          eid = {arXiv:2010.11929},
        pages = {arXiv:2010.11929},
          doi = {10.48550/arXiv.2010.11929},
archivePrefix = {arXiv},
       eprint = {2010.11929},
 primaryClass = {cs.CV},
       adsurl = {https://ui.adsabs.harvard.edu/abs/2020arXiv201011929D}
}

@ARTICLE{transformer_vaswani,
       author = {{Vaswani}, Ashish and {Shazeer}, Noam and {Parmar}, Niki and {Uszkoreit}, Jakob and {Jones}, Llion and {Gomez}, Aidan N. and {Kaiser}, Lukasz and {Polosukhin}, Illia},
        title = "{Attention Is All You Need}",
      journal = {arXiv e-prints},
         year = 2017,
        month = jun,
          eid = {arXiv:1706.03762},
        pages = {arXiv:1706.03762},
          doi = {10.48550/arXiv.1706.03762},
archivePrefix = {arXiv},
       eprint = {1706.03762},
 primaryClass = {cs.CL},
       adsurl = {https://ui.adsabs.harvard.edu/abs/2017arXiv170603762V}
}

@ARTICLE{2013A&A...558A..33A,
       author = {{Astropy Collaboration} and {Robitaille}, Thomas P. and
         {Tollerud}, Erik J. and {Greenfield}, Perry and {Droettboom}, Michael and
         {Bray}, Erik and {Aldcroft}, Tom and {Davis}, Matt and
         {Ginsburg}, Adam and {Price-Whelan}, Adrian M. and
         {Kerzendorf}, Wolfgang E. and {Conley}, Alexander and {Crighton}, Neil and
         {Barbary}, Kyle and {Muna}, Demitri and {Ferguson}, Henry and
         {Grollier}, Fr{\'e}d{\'e}ric and {Parikh}, Madhura M. and
         {Nair}, Prasanth H. and {Unther}, Hans M. and {Deil}, Christoph and
         {Woillez}, Julien and {Conseil}, Simon and {Kramer}, Roban and
         {Turner}, James E.~H. and {Singer}, Leo and {Fox}, Ryan and
         {Weaver}, Benjamin A. and {Zabalza}, Victor and {Edwards}, Zachary I. and
         {Azalee Bostroem}, K. and {Burke}, D.~J. and {Casey}, Andrew R. and
         {Crawford}, Steven M. and {Dencheva}, Nadia and {Ely}, Justin and
         {Jenness}, Tim and {Labrie}, Kathleen and {Lim}, Pey Lian and
         {Pierfederici}, Francesco and {Pontzen}, Andrew and {Ptak}, Andy and
         {Refsdal}, Brian and {Servillat}, Mathieu and {Streicher}, Ole},
        title = "{Astropy: A community Python package for astronomy}",
      journal = {\aap},
         year = "2013",
        month = "Oct",
       volume = {558},
          eid = {A33},
        pages = {A33},
          doi = {10.1051/0004-6361/201322068},
archivePrefix = {arXiv},
       eprint = {1307.6212},
 primaryClass = {astro-ph.IM},
       adsurl = {https://ui.adsabs.harvard.edu/abs/2013A&A...558A..33A}
}
\bibliographystyle{aasjournalv7}



\end{document}